\documentstyle[epsf]{article}

\newcommand{\CXgqbLali}{G^{Lb}_{\tilde{u}_\alpha \tilde{\chi}^-_i}}
\newcommand{\CXgqbRali}{G^{Rb}_{\tilde{u}_\alpha \tilde{\chi}^-_i}}
\newcommand{\CXgqbLbej}{G^{Lb}_{\tilde{u}_\beta \tilde{\chi}^-_j}}
\newcommand{\CXgqbRbej}{G^{Rb}_{\tilde{u}_\beta \tilde{\chi}^-_j}}
\newcommand{\CXgqsLali}{G^{Ls\ast}_{\tilde{u}_\alpha \tilde{\chi}^-_i}}

\newcommand{\CXgqsLbej}{G^{Ls\ast}_{\tilde{u}_\beta \tilde{\chi}^-_j}}

\newcommand{\mcxi}{m_{\tilde{\chi}^-_i}}
\newcommand{\mcxj}{m_{\tilde{\chi}^-_j}}
\newcommand{\sqalcxi}{x_{\tilde{u}_\alpha \tilde{\chi}^-_i}}
\newcommand{\sqalcxj}{x_{\tilde{u}_\alpha \tilde{\chi}^-_j}}
\newcommand{\sqbecxj}{x_{\tilde{u}_\beta \tilde{\chi}^-_j}}
\newcommand{\cxicxj}{x_{\tilde{\chi}^-_i \tilde{\chi}^-_j}}

\newcommand{\CXgnuRstari}{G^{Rl\ast}_{\tilde{\chi}^-_i \tilde{\nu}_l}}
\newcommand{\CXgnuLstari}{G^{Ll\ast}_{\tilde{\chi}^-_i \tilde{\nu}_l}}
\newcommand{\CXgnuRj}{G^{Rl}_{\tilde{\nu}_l \tilde{\chi}^-_j}}
\newcommand{\CXgnuLj}{G^{Ll}_{\tilde{\tilde{\nu}_l \chi}^-_j}}
\newcommand{\snucxj}{x_{\tilde{\nu}_l \tilde{\chi}^-_j}}

\newcommand{\NXgqbLali}{G^{Lb}_{\tilde{d}_\alpha \tilde{\chi}^0_i}}
\newcommand{\NXgqbRali}{G^{Rb}_{\tilde{d}_\alpha \tilde{\chi}^0_i}}
\newcommand{\NXgqbLbej}{G^{Lb}_{\tilde{d}_\beta \tilde{\chi}^0_j}}
\newcommand{\NXgqbRbej}{G^{Rb}_{\tilde{d}_\beta \tilde{\chi}^0_j}}
\newcommand{\NXgqsLali}{G^{Ls\ast}_{\tilde{d}_\alpha \tilde{\chi}^0_i}}
\newcommand{\NXgqsRali}{G^{Rs\ast}_{\tilde{d}_\alpha \tilde{\chi}^0_i}}
\newcommand{\NXgqsLbej}{G^{Ls}_{\tilde{d}_\beta \tilde{\chi}^0_j}}
\newcommand{\NXgqsRbej}{G^{Rs}_{\tilde{d}_\beta \tilde{\chi}^0_j}}
\newcommand{\mnxi}{m_{\tilde{\chi}^0_i}}
\newcommand{\mnxj}{m_{\tilde{\chi}^0_j}}
\newcommand{\sqalnxi}{x_{\tilde{d}_\alpha \tilde{\chi}^0_i}}
\newcommand{\sqalnxj}{x_{\tilde{d}_\alpha \tilde{\chi}^0_j}}
\newcommand{\sqbenxj}{x_{\tilde{d}_\beta \tilde{\chi}^0_j}}
\newcommand{\nxinxj}{x_{\tilde{\chi}^0_i \tilde{\chi}^0_j}}

\newcommand{\NXzggLij}{G^{Lz}_{\tilde{\chi}^0_i \tilde{\chi}^0_j}}
\newcommand{\NXzggRij}{G^{Rz}_{\tilde{\chi}^0_i \tilde{\chi}^0_j}}
\newcommand{\NXglRstarim}{G^{Rl\ast}_{\tilde{l}_m \tilde{\chi}^0_i}}
\newcommand{\NXglLstarim}{G^{Ll\ast}_{\tilde{l}_m \tilde{\chi}^0_i}}
\newcommand{\NXglRstarjm}{G^{Rl\ast}_{\tilde{l}_m \tilde{\chi}^0_j}}
\newcommand{\NXglLstarjm}{G^{Ll\ast}_{\tilde{l}_m \tilde{\chi}^0_j}}
\newcommand{\NXglRim}{G^{Rl}_{\tilde{l}_m \tilde{\chi}^0_i}}
\newcommand{\NXglLim}{G^{Ll}_{\tilde{l}_m \tilde{\chi}^0_i}}
\newcommand{\NXglRjm}{G^{Rl}_{\tilde{l}_m \tilde{\chi}^0_j}}
\newcommand{\NXglLjm}{G^{Ll}_{\tilde{l}_m \tilde{\chi}^0_j}}
\newcommand{\slmj}{x_{\tilde{l}_m \tilde{\chi}^0_j}}

\newcommand{\GGgqbLal}{G^{Lb}_{\tilde{d}_\alpha \tilde{g}}}
\newcommand{\GGgqbRal}{G^{Rb}_{\tilde{d}_\alpha \tilde{g}}}
\newcommand{\GGgqbLbe}{G^{Lb}_{\tilde{d}_\beta \tilde{g}}}
\newcommand{\GGgqbRbe}{G^{Rb}_{\tilde{d}_\beta \tilde{g}}}
\newcommand{\GGgqsLal}{G^{Ls\ast}_{\tilde{d}_\alpha \tilde{g}}}
\newcommand{\GGgqsRal}{G^{Rs\ast}_{\tilde{d}_\alpha \tilde{g}}}
\newcommand{\GGgqsLbe}{G^{Ls\ast}_{\tilde{d}_\beta \tilde{g}}}
\newcommand{\GGgqsRbe}{G^{Rs\ast}_{\tilde{d}_\beta \tilde{g}}}
\newcommand{\mgg}{m_{\tilde{g}}}
\newcommand{\sqalgg}{x_{\tilde{d}_\alpha \tilde{g}}}
\newcommand{\sqbegg}{x_{\tilde{d}_\beta \tilde{g}}}

\newcommand{\shHualbe}{G^{H^0}_{\tilde{u}_\alpha \tilde{u}_\beta}}
\newcommand{\shhualbe}{G^{h^0}_{\tilde{u}_\alpha \tilde{u}_\beta}}
\newcommand{\shGualbe}{G^{G^0}_{\tilde{u}_\alpha \tilde{u}_\beta}}
\newcommand{\shAualbe}{G^{A^0}_{\tilde{u}_\alpha \tilde{u}_\beta}}
\newcommand{\shHdalbe}{G^{H^0}_{\tilde{d}_\alpha \tilde{d}_\beta}}
\newcommand{\shhdalbe}{G^{h^0}_{\tilde{d}_\alpha \tilde{d}_\beta}}
\newcommand{\shGdalbe}{G^{G^0}_{\tilde{d}_\alpha \tilde{d}_\beta}}
\newcommand{\shAdalbe}{G^{A^0}_{\tilde{d}_\alpha \tilde{d}_\beta}}
\newcommand{\gcxHLij}{G^{LH^0}_{\tilde{\chi}^-_i \tilde{\chi}^-_j}}
\newcommand{\gcxHRij}{G^{RH^0}_{\tilde{\chi}^-_i \tilde{\chi}^-_j}}
\newcommand{\gcxhLij}{G^{Lh^0}_{\tilde{\chi}^-_i \tilde{\chi}^-_j}}
\newcommand{\gcxhRij}{G^{Rh^0}_{\tilde{\chi}^-_i \tilde{\chi}^-_j}}
\newcommand{\gcxGLij}{G^{LG^0}_{\tilde{\chi}^-_i \tilde{\chi}^-_j}}
\newcommand{\gcxGRij}{G^{RG^0}_{\tilde{\chi}^-_i \tilde{\chi}^-_j}}
\newcommand{\gcxALij}{G^{LA^0}_{\tilde{\chi}^-_i \tilde{\chi}^-_j}}
\newcommand{\gcxARij}{G^{RA^0}_{\tilde{\chi}^-_i \tilde{\chi}^-_j}}
\newcommand{\gnxHLij}{G^{LH^0}_{\tilde{\chi}^0_i \tilde{\chi}^0_j}}
\newcommand{\gnxHRij}{G^{RH^0}_{\tilde{\chi}^0_i \tilde{\chi}^0_j}}
\newcommand{\gnxhLij}{G^{Lh^0}_{\tilde{\chi}^0_i \tilde{\chi}^0_j}}
\newcommand{\gnxhRij}{G^{Rh^0}_{\tilde{\chi}^0_i \tilde{\chi}^0_j}}
\newcommand{\gnxGLij}{G^{LG^0}_{\tilde{\chi}^0_i \tilde{\chi}^0_j}}
\newcommand{\gnxGRij}{G^{RG^0}_{\tilde{\chi}^0_i \tilde{\chi}^0_j}}
\newcommand{\gnxALij}{G^{LA^0}_{\tilde{\chi}^0_i \tilde{\chi}^0_j}}
\newcommand{\gnxARij}{G^{RA^0}_{\tilde{\chi}^0_i \tilde{\chi}^0_j}}
\newcommand{\slmnxj}{x_{\tilde{l}_m \tilde{\chi}^0_j}}

\def \M{{\cal M}}

\begin{document}

\title{$B_s \rightarrow \mu^+ \mu^-$ and $B \rightarrow X_s \mu^+ \mu^-$
 in MSSM}

\author{
Chao-Shang Huang\footnote{Email: csh@itp.ac.cn}\\
Institute of Theoretical Physics, Chinese Academy of Sciences,\\
P. O. Box 2735, Beijing 100080, P. R. China\\
and\\
Xiao-Hong Wu\footnote{Email: wuxh@th.phy.pku.cn}\\
CCAST (World Lab.), P. O. Box 8730, Beijing 100080, P. R. China\\
Institute of Theoretical Physics, Chinese Academy of Sciences,\\
P. O. Box 2735, Beijing 100080, P. R. China
}

\date{March 6, 2003}

\maketitle

\begin{abstract}
Pure leptonic and semileptonic rare B decays, $B_s \rightarrow \mu^+ \mu^-$
and $B \rightarrow X_s \mu^+ \mu^-$
in the minimal supersymmetric Standard Model (MSSM),
in particular, gluino and neutralino contributions to
the decays, are discussed under Br($B \rightarrow X_s \gamma$)
and other experimental constraints .
The general scalar quark mass matrices as the new sources of
flavor violation are considered.
We present Wilson coefficients of $bs$ transitions
from $\gamma$, $Z$, and neutral Higgs boson
penguin diagrams by using vertex mixing method to deal with
scalar down-type quark flavor changing  and also give
their expressions in MIA to show different sources of enhancements.
We find that under the experimental constraints,
with large mixing of left-handed and right-handed sbottom,
$C^{(\prime)}_{10}$ can be enhanced by $10\%$ compared with SM,
in two cases, heavy gluino
and fine-tuning between $\delta^{dLL}_{23}$($\delta^{dRR}_{23}$)
and $\delta^{dLR}_{23}$($\delta^{dRL}_{23}$) terms in $C^{(\prime)}_7$.
Particularly, $C_{10}$ and $C_{10}^{\prime}$ can reach a $20\%$ enhancement in
some regions of parameters  under experimental constraints.
When CP-odd Higgs $A^0$ is not too heavy ($\sim 250$GeV),
and $\tan\beta$ is large ($\sim 40$), neutral Higgs boson penguins
with gluino and down-type squark in the loop can significantly
contribute to the $bs$ transition and the contributions can
compete with those due to the chargino and up-type squark loop.
\end{abstract}


\newpage
\section{Introduction}\label{sect:intro}
Pure leptonic and semileptonic flavor changing neutral current
(FCNC) rare B decays,
\begin{equation}
B_s \rightarrow \mu^+ \mu^-,\hspace{5mm} B \rightarrow X_s \mu^+ \mu^-
\label{bs}
\end{equation}
have received much attention recent years due to clear backgrounds
and ongoing experiments at BaBar~\cite{babarbook} and BELLE and
forthcoming projects at Tevatron~\cite{btev} and LHC~\cite{lhcb} as
well as sensitivity to models beyond the Standard Model (SM). The
current experimental result of Br($B \rightarrow X_s \mu^+ \mu^-$)
by the BELLE collaboration~\cite{BELLEbsll} is
\begin{equation}\label{semi}
{\rm Br}(B \rightarrow X_s \mu^+ \mu^-)_{\rm exp} = 7.9 \pm 2.1^{+2.1}_{-1.5}
\times 10^{-6}
\end{equation}
and CDF~\cite{CDFbsmu} upper limit on Br($B_s \rightarrow \mu^+ \mu^-$) is
\begin{equation}
{\rm Br}(B_s \rightarrow \mu^+ \mu^-)_{\rm exp} < 2.6 \times 10^{-6}
{\rm\hspace{2mm} at\hspace{2mm} 90\% C.L.}
\end{equation}

In the SM,  these processes vanish at tree level,
while they occur at one-loop level with the charged gauge boson $W^\pm$
and up-type quarks in the loop.
In the minimal supersymmetric Standard Model (MSSM),
there are five kinds of contributions to partonic level process
$b \rightarrow s \mu^+ \mu^-$ at one-loop level,
depending on specific particles propagated in the loop,
($1$) Standard Model gauge boson $W^\pm$ and up-type quarks (SM contribution);
($2$) charged Higgs $H^\pm$ and up-type quarks (charged Higgs contribution);
($3$) chargino and scalar up-type quarks (chargino contribution);
($4$) neutralino and scalar down-type quarks (neutralino contribution);
($5$) gluino and scalar down-type quarks (gluino contribution).
The flavor structure of the sfermion sector in MSSM depends on
the soft terms which are determined by the supersymmetry breaking mechanism,
in addition to the superpotential.
In the minimal flavor violation (MFV) scenarios of MSSM, squarks
are assumed to rotate in flavor
bases aligned with the corresponding quark sector and
the only source of flavor violation is the
usual Cabibbo-Kobayashi-Maskawa (CKM) matrix in SM. In some MFV scenarios
such as the constrained MSSM (mSUGRA, string-inspired flipped SU(5), etc.)
and gauge mediation supersymmetry breaking (GMSB)
where the soft terms at some high scale
(the grand unification scale or Plank scale or messenger scale) are
characterized by the universality of sfermion masses and
the proportionality of the trilinear terms, the flavor violation
in sfermion sector at the electroweak (EW) scale is generated
radiatively and consequently, in general, small
\footnote{In some large $\tan\beta$ regions of the parameter space
it becomes non-negligible ~\cite{bor}.}. Therefore,
comparing with the first three kinds of contributions, the
last two kinds of contributions, i. e., neutralino and gluino contributions,
are negligible~\cite{bsll}.

There are new sources of flavor violation in MSSM. Besides the CKM matrix, the
the $6\times 6$ squark mass matrices are generally not diagonal
in flavor (generation) indices in the super-CKM basis
in which superfields are rotated in such a way that
the mass matrices of the quark field components of
the superfields are diagonal.
This rotation non-alignment in the quark and squark sectors can induce
large flavor off-diagonal couplings such as the coupling of gluino to
the quark and squark which belong to different generations.
There exist two different kinds of methods to deal with
flavor changing vertices induced by  flavor mixing in the squark
mass matrices in the literature~\cite{randall}.
One works in quark and squark mass eigenstates with induced flavor changing
couplings, so called "vertex mixing" (VM).
The other method, "mass insertion approximation"(MIA)~\cite{mi},
works in flavor diagonal gaugino couplings $\tilde{g}q\tilde{q}$ and
diagonal quark mass matrices with all the flavor changes rested on the
off-diagonal sfermion propagators.
The MIA can be obtained in VM through Taylor expansion of
nearly degenerate squark masses $m_{\tilde{q}_i}$ around the common squark mass
$m_{\tilde{q}}$, $m^2_{\tilde{q}_i} \simeq m^2_{\tilde{q}} (1 + \Delta_i)$.
Thus MIA can work well for nearly degenerate squark masses and, in general,
its reliability can be checked only a posteriori.
However, for its simplicity, it has been widely used as a model independent
analysis to find the constraints on different off-diagonal parts of
squark mass matrices from experiments~\cite{gabbiani}.
It is clear that VM remains valid even when flavor
off-diagonal squark mass matrix elements are large and
there is no approximation which has been assumed.

Because the experiment of Br($B \rightarrow X_s \gamma$) only constrains
$|C_7(m_b)|^2 + |C^{\prime}_7(m_b)|^2$, with the overall sign flip of
$C_7(m_b)$ from MSSM {\it vs} SM,
the branch ratio of $B \rightarrow X_s l^+ l^-$ can
be enhanced from the term Re($C_7(m_b) C^{eff\ast}_9(m_b)$)
in the constrained MSSM with gluino and neutralino contributions
neglected~\cite{c7sign,bsll,goto97}, while in these analysis
supersymmetry contributions to $C_9$ and $C_{10}$ are at most
changed by $\pm 5\%$ compared with the SM values.
The chargino contribution in the extended MFV model is analyzed in
ref~\cite{ali01}. $B \rightarrow X_s l^+ l^-$ has been analyzed in
the left-right supersymmetric model recently~\cite{frank02}.
In MSSM, gluino induced FCNC process
$b \rightarrow s l^+ l^-$ is studied
in ref~\cite{kim99}, and chargino, gluino induced effects are studied
in ref~\cite{lunghi00} in MIA. In these works the contributions from exchanging
neutral Higgs bosons (NHBs) are not included. However, for $l = \mu, \tau$,
when $\tan\beta$ is large and $M_A^0$ is not too large (say, 250 GeV),
the NHBs contributions
can become significant due to the $\tan^3\beta$ enhancement of
the corresponding Wilson coefficients in some regions of
the parameter space~\cite{huang,babu} and the NHBs contributions
to $B\rightarrow X_s l^+l^-$ and $B_q \rightarrow l^+ l^-$
($q = d, s$, $l = \mu, \tau$) in the
constrained MSSM have been investigated in refs.~\cite{huang,babu,bsll1,bll}.
Using the VM method, the gluino and neutralino induced FCNC processes
$b \rightarrow s \mu^+ \mu^-$ ~\cite{xiong} and
$B_q \rightarrow \mu^+ \mu^-$ ~\cite{urban} have recently been analyzed,
including the NHBs contributions, in the constrained MSSM and in MFV models
respectively.
In SUSY models with non-minimal sources of flavor mixing,
the constraints on different flavor violation parameters from
Br($B \rightarrow X_s \gamma$) have been considered
in ref.~\cite{bghw,everett01,besmer},
and $B_{s,d} \rightarrow l^+ l^-$ at large $\tan\beta$ has been
investigated in ref.~\cite{isidori02}.

It is well-known that the effects of the primed counterparts of
usual operators are suppressed by $\frac{m_s}{m_b}$ and
consequently negligible in SM because they have the opposite chiralities.
In MFV models their effects are also negligible, as shown in ref.~\cite{urban}.
However, in MSSM their effects can be significant,
since the flavor non-diagonal squark mass matrix elements are free parameters.
Part of the primed counterparts of usual operators relevant to B rare
leptonic and semileptonic decays have been considered in ref.\cite{lunghi00}.

In this paper, we extend our previous analyses to include gluino and
neutralino contributions and all operators responsible for
B rare leptonic and semileptonic decays in MSSM. We calculate
the Wilson coefficients using the VM method and also give
their expressions in MIA to show different sources of enhancements.
In numerical analyses we take into account
constraints from Br($B \rightarrow X_s \gamma$), $\Delta M_{B_s}$ and
the lower bounds of superpartner masses and Higgs masses as well as
$B\rightarrow X_s g$ and hadronic charmless $B$ decays.
We have carefully analyzed different sources of
enhancements of $C^{(\prime)}_{10}$ (arising from the $Z$ penguin)
and $C^{(\prime)}_{Q_{1,2}}$ (arising from the NHB penguins),
related to the general scalar-down quark mass matrix.
We find that under the experimental constraints,
with large mixing of left-handed and right-handed sbottom,
$C^{(\prime)}_{10}$ can be enhanced by about $10\%$ compared to the
Standard Model in two cases, heavy gluino
and fine-tuning between $\delta^{dLL}_{23}$($\delta^{dRR}_{23}$)
and $\delta^{dLR}_{23}$($\delta^{dRL}_{23}$) terms in $C^{(\prime)}_7$.
In particular, $C_{10}$ and $C_{10}^{\prime}$ can reach a $20\%$ enhancement in
some regions of parameters  under experimental constraints.
When CP-odd Higgs $A^0$ is not too heavy ($\sim 250$GeV),
and $\tan\beta$ is large ($\sim 40$), neutral Higgs boson penguins with
gluino and down-type squark in the loop can significantly
contribute to the $bs$ transition and the contributions can compete with
those due to the chargino and scalar up-type quark loop.

The paper is arranged as following. In section \ref{sect:hamiltonian},
we define our notations and
consider the effective Hamiltonian and branching ratios of pure leptonic
and semileptonic
rare B decays. In section \ref{sect:squark} we briefly recall the squark mass
matrices and discuss the choice of parameters. In section \ref{sect:numerical},
we present our numerical analysis on the possible enhancement of
$C_{10}^{(\prime)}$ and $C_{Q_{1,2}}^{(\prime)}$
in the case of switching on only the gluino
(or neutralino) and SM contributions.
We search for maximums of $C_{10}$ and
$C_{10}^{\prime}$ under experimental constraints,
switching on all the contributions.  Section 5 is devoted to
give the numerical results for $B\rightarrow X_s l^+l^-$
and $B_s\rightarrow l^+l^-$.
Conclusions are drawn in section \ref{sect:conclusion}.
In the Appendix, Wilson coefficients at $m_w$ scale are given.

\section{Effective Hamiltonian}\label{sect:hamiltonian}
The effective Hamiltonian for $B \rightarrow X_s l^+ l^-$
and $B_s\rightarrow l^+l^-$ can be written as
\begin{eqnarray}
{\cal H}_{eff} = \frac{4 G_F}{\sqrt{2}} \lambda_t \left [ \sum_{i=1}^{6}
C_i(\mu) O_i(\mu)
+ \sum_{i=7}^{10}(C_i(\mu)O_i(\mu)+C_i^{\prime}(\mu)O_i^{\prime}(\mu))
+ \sum_{i=1}^{10}
(C_{Q_i}(\mu) Q_i(\mu)+C_{Q_i}^{\prime}(\mu) Q_i^{\prime}(\mu)) \right ]
\end{eqnarray}
where $\lambda_t = V_{tb} V_{ts}^\ast$, $O_i$ and $Q_i$ (i=1,...,10)
can be found in ref.~\cite{operator} and
~\cite{dhh} respectively, and the primed operators,
the counterpart of the unprimed operators, are obtained by replacing
the chiralities in the corresponding unprimed operators with
opposite ones. The explicit expressions of the operators governing
$B \rightarrow X_s l^+ l^-$ and $B_s\rightarrow l^+l^-$ are given by
\begin{eqnarray}
O_7 &=& \frac{e}{16\pi^2}
m_b (\bar{s} \sigma_{\mu\nu} P_R b)F^{\mu\nu}, \hspace{5mm}
O_7^\prime = \frac{e}{16\pi^2}
m_b (\bar{s} \sigma_{\mu\nu} P_L b)F^{\mu\nu}\nonumber\\
O_9 &=& \frac{e^2}{16\pi^2}
(\bar{s}\gamma_\mu P_L b) (\bar{l}\gamma^\mu l), \hspace{5mm}
O_9^\prime = \frac{e^2}{16\pi^2}
(\bar{s}\gamma_\mu P_R b) (\bar{l}\gamma^\mu l)\nonumber\\
O_{10} &=& \frac{e^2}{16\pi^2}
(\bar{s}\gamma_\mu P_L b) (\bar{l}\gamma^\mu \gamma_5 l), \hspace{5mm}
O_{10}^\prime = \frac{e^2}{16\pi^2}
(\bar{s}\gamma_\mu P_R b)
(\bar{l}\gamma^\mu \gamma_5 l)\nonumber\\
Q_1 &=& \frac{e^2}{16\pi^2}
(\bar{s} P_R b)(\bar{l} l), \hspace{5mm}
Q_1^\prime = \frac{e^2}{16\pi^2}
(\bar{s} P_L b)(\bar{l} l)\nonumber\\
Q_2 &=& \frac{e^2}{16\pi^2}
(\bar{s} P_R b)(\bar{l} \gamma_5 l), \hspace{5mm}
Q_2^\prime = \frac{e^2}{16\pi^2}
(\bar{s} P_L b)(\bar{l} \gamma_5 l)
\end{eqnarray}
where $P_L \equiv (1 - \gamma_5)/2$ and
$P_R \equiv (1 + \gamma_5)/2$\footnote{Note that
$Q_{1,2}=\frac{1}{m_b}\frac{e^2}{16\pi^2} O_{S,P}$ and $O_{S,P}$
are used in some papers (see, e.g., ref.\cite{ln}).}.
We also consider the operators
\begin{eqnarray}
O_8 &=& \frac{g_s}{16\pi^2}
m_b (\bar{s}_\alpha T^a_{\alpha\beta}\sigma_{\mu\nu}
P_R b_\beta)G^{a\mu\nu}, \hspace{5mm}
O_8^\prime = \frac{g_s}{16\pi^2}
m_b (\bar{s}_\alpha T^a_{\alpha\beta}\sigma_{\mu\nu}
P_L b_\beta)G^{a\mu\nu}
\end{eqnarray}
in order to include the constraints from
$B \rightarrow X_s g$ and hadronic charmless B decays into the analysis.
In SM the Wilson coefficients of the prime operaters are suppressed
by $\frac{m_s}{m_b}$ with respect to those of unprimed operators
and the statement is also true in MFV scenarios of MSSM~\cite{urban}.
However, in MSSM, the statement is, in general, not valid
due to the presence of new sources of flavor violation.
The running of Wilson coefficients $C_i$ and $C_{Q_i}$ from $m_w$ to
$m_b$ in the leading order approximation (LO) is given
in refs~\cite{operator} and ~\cite{dhh} respectively.
The evolution of part of the primed operators has been
given in ref.~\cite{bghw}.  Although the mixing between $O_i$ in the
next-to-leading order (NLO)
has been studied, the mixing of $O_i$ with $Q_i$ in NLO has not been given.
So we shall use only the LO results for consistence.
We present $m_w$ scale Wilson coefficients
$C^{(\prime)}_i$ and $C^{(\prime)}_{Q_i}$ in Appendix.
In order to see the dependences of
Wilson coefficients on new flavor violation parameters,
as an illustration, we present all the relevant Wilson coefficients
at the $m_b$ scale induced by gluino in MIA as follows
\begin{eqnarray}
\label{c7}
C^{\tilde{g} (\prime)}_7(m_b) &=& 18.7 \frac{400{\rm GeV}}{m_{\tilde{g}}}
\delta^{dLL(RR)}_{23} \delta^{dLR}_{33} -
34.6 \frac{400{\rm GeV}}{m_{\tilde{g}}} \delta^{dLR(RL)}_{23} + \nonumber\\
&& 0.07 (\frac{400{\rm GeV}}{m_{\tilde{g}}})^2 \delta^{dLL(RR)}_{23} -
0.04 (\frac{400{\rm GeV}}{m_{\tilde{g}}})^2 \delta^{dLR(RL)}_{23}
\delta^{dLR}_{33}\\
\label{c8}
C^{\tilde{g} (\prime)}_8(m_b) &=& 38.5 \frac{400{\rm GeV}}{m_{\tilde{g}}}
\delta^{dLL(RR)}_{23} \delta^{dLR}_{23} -
92.9 \frac{400{\rm GeV}}{m_{\tilde{g}}} \delta^{dLR(RL)}_{23} + \nonumber\\
&& 0.15 (\frac{400{\rm GeV}}{m_{\tilde{g}}})^2 \delta^{dLL(RR)}_{23} -
0.08 (\frac{400{\rm GeV}}{m_{\tilde{g}}})^2 \delta^{dLR(RL)}_{23}
\delta^{dLR}_{33}\\
\label{c9}
C^{\tilde{g} (\prime)}_9(m_b) &=&
- 0.21 (\frac{400{\rm GeV}}{m_{\tilde{g}}})^2 \delta^{dLL(RR)}_{23}
+ 0.17 (\frac{400{\rm GeV}}{m_{\tilde{g}}})^2 \delta^{dLR(RL)}_{23}
\delta^{dLR}_{33} +\nonumber\\
&& 0.67 \delta^{dLL(RR)}_{23} (\delta^{dLR}_{33})^2 -
1.68 \delta^{dLR(RL)}_{23} \delta^{dLR}_{33}\\
\label{c10}
C^{\tilde{g} (\prime)}_{10}(m_b) &=& -
8.37 \delta^{dLL(RR)}_{23} (\delta^{dLR}_{33})^2 +
21.0 \delta^{dLR(RL)}_{23} \delta^{dLR}_{33} \\
\label{cs}
C^{\tilde{g} (\prime)}_{Q_1}(m_b) &=& -
4.61 (\frac{200{\rm GeV}}{m_{A^0}})^2 (\frac{\tan\beta}{40})^2
\frac{m_{\tilde{g}}}{400{\rm GeV}} \delta^{dLL(RR)}_{23}
\delta^{dLR}_{33} \\
\label{cp}
C^{\tilde{g} (\prime)}_{Q_2}(m_b) &=& \mp C^{\tilde{g} (\prime)}_{Q_1}(m_b)
\end{eqnarray}
Where $\delta^{qAB}_{ij} \equiv \frac{ (\M^2_{\tilde{q}AB})_{ij} -
\tilde{m}^2 \delta_{ij} \delta_{AB} }{\tilde{m}^2}$ ($M^2_{\tilde{q}AB}$
is the scalar quark mass matrix, $q = u, d$, $A, B = L, R$, $i, j = 1, 2, 3$,
$\tilde{m}$ is the common scalar quark mass),
the one-loop functions (which are given in appendix)
at $x = \frac{m^2_{\tilde{q}}}{m^2_{\tilde{g}}} = 1$ have been used,
$\tan\beta$ is the ratio of two Higgs vacuum expectation value $v_U$
and $v_D$, $\tan\beta = \frac{v_U}{v_D}$, and $m_{H^0}=m_{A^0}$,
which is a good approximation in the case of $m_{A^0}^2/m_{Z}^2 \gg 1$,
has been assumed.
Here we have expanded the Wilson coefficients to the double MI,
as investigated in ref~\cite{everett01},
which is non-neglectable if the mixing between
left-handed and right-handed sbottoms  is large.
In eqs.(\ref{c7}) and (\ref{c8}) the first two terms are
suppressed by $\frac{1}{m_{\tilde{g}}}$
(as can be seen from Appendix, although they are enhanced by
a factor $\frac{m_{\tilde{g}}}{m_b}$,
they suffer from the $\frac{m^2_w}{m^2_{\tilde{g}}}$ suppression).
In eq.(\ref{c9}) for $C^{(\prime)}_9(m_b)$,
the first two terms come from the $\gamma$ penguin,
which is suppressed by $\frac{m_w^2}{m^2_{\tilde{g}}}$.
But eq.(\ref{c10}) for $C^{(\prime)}_{10}(m_b)$ and
the last two terms in eq.(\ref{c9}) from the $Z$ penguin is not suppressed,
which is noticed as non-decoupling of the $bsZ$ coupling
in ref.~\cite{buchalla00}.
Therefore, there is a possibility that $C_{10}$ can be enhanced even
if the gluino is heavy.
In eq.(\ref{cs}) and eq.(\ref{cp}) for $C^{(\prime)}_{Q_{1,2}}$,
we present $\delta^{dLL(RR)}_{23}$ contribution in the MIA,
which is enhanced by a factor $\frac{m_{\tilde{g}}}{m_b}$\footnote{The
$\delta^{dLR(RL)}_{23}$ contributions to $C^{(\prime)}_{Q_{1,2}}$
from self-energy type and penguin diagrams are cancelled provided
that $m_{A^0}=m_{H^0}$ and MIA is valid.}.
Compared to the chargino contribution,
the gluino contribution is the same important provided
that $m_b\tan\beta\sim m_{\tilde{g}}$.
We can read that $C^{(\prime)}_{Q_{1,2}}$
can be large if CP-odd Higgs $A^0$ is not too heavy
(say, $\leq 400$ GeV) and $\tan\beta$ is large.

The leading order $B_s \rightarrow X_s \gamma$ branching ratio
normalized to ${\rm Br}(B \rightarrow X_c e \bar{\nu})$ is given as
\begin{equation}
{\rm Br}(B \rightarrow X_s \gamma) = \frac{6 \alpha_{\rm em}}{\pi f(z)}
|\frac{V_{tb} V^\ast_{ts}}{V{cb}}|^2 {\rm Br}(B \rightarrow X_c e \bar{\nu})
(|C_7(m_b)|^2 + |C^{\prime}_7(m_b)|^2),
\end{equation}
where $\sqrt{z} = m_c^{\rm pole}/m_b^{\rm pole}$,
$f(z)$ is the phase space function.

The branching ratio (Br) $B \rightarrow X_s \mu^+ \mu^-$ normalized to
${\rm Br}(B \rightarrow X_c e \bar{\nu})$ is given as
\begin{eqnarray}\label{slbr}
\frac{d{\rm Br}(B \rightarrow X_s \mu^+ \mu^-)}{ds} &=& {\rm Br}(B
\rightarrow X_c e \bar{\nu}) \frac{\alpha_{\rm em}^2}{4 \pi^2 f(z)} (1
- s)^2 \sqrt{1 - \frac{4 t^2}{s}} \frac{|V_{tb}
V_{ts}^\ast|^2}{|V_{cb}|^2} D(s)\;, \nonumber\\ D(s) &=&
|C_9^{\rm eff}(m_b)|^2 (1 + \frac{2 t^2}{s})(1 + 2s) +
4 |C_7(m_b)|^2 (1 + \frac{2 t^2}{s})(1 + \frac{2}{s}) + \nonumber\\
&& |C_{10}(m_b)|^2[(1 + 2s) + \frac{2 t^2}{s}(1 - 4s)] +
12 {\rm Re}(C_7(m_b) C_9^{\rm eff*}(m_b))(1 + \frac{2 t^2}{s}) + \nonumber\\
&& \frac{3}{2}|C_{Q_1}(m_b)|^2 (s - 4 t^2) + \frac{3}{2}|C_{Q_2}(m_b)|^2 s +
6 {\rm Re}(C_{10}(m_b) C^*_{Q_2}(m_b))t + \nonumber\\
&& ( C_i(m_b) \leftrightarrow C^\prime_i(m_b) ).
\end{eqnarray}
where $t=\frac{m_\mu}{m_b}$ and $s = \frac{M^2_{\mu^+\mu^-}}{M^2_B}$.
In SM, $C_7(m_b)$ and $C_9^{\rm eff*}(m_b)$ have opposite sign, so the term
Re$(C_7(m_b) C_9^{\rm eff*}(m_b)))$ decreases the overall results above.
In MSSM, there exist some parameter regions, where supersymmetry contributions
to $C_7$ can have opposite sign compared with $C^{\rm SM}_7(m_b)$, and even
make the sign of $C_7(m_b)$ opposite to $C^{\rm SM}_7$
but its size approximately equal to $C^{\rm SM}_7(m_b)$. In this case the term
Re$(C_7(m_b) C_9^{\rm eff*}(m_b)))$ adds constructively to
$\frac{d{\rm Br}(B \rightarrow X_s \mu^+ \mu^-)}{ds}$, which can make the
Br of $B \rightarrow X_s l^+ l^-$ enhanced by
approximately $50\%$ compared to that in SM~\cite{goto97}.
From the formula above, we can see that with large $C^{(\prime)}_{Q_{1,2}}$,
$B \rightarrow X_s \mu^+ \mu^-$ can be enhanced greatly.

The branching ratio $B_s \rightarrow \mu^+ \mu^-$is given as
\begin{eqnarray}
{\rm Br}(B_s \rightarrow \mu^+ \mu^-) &=&
\frac{G_F^2 \alpha^2_{\rm em}}{64 \pi^3}
m^3_{B_s} \tau_{B_s} f^2_{B_s} |\lambda_t|^2 \sqrt{1 - 4 \hat{m}^2}
[(1 - 4\hat{m}^2) |C_{Q_1}(m_b) - C^\prime_{Q_1}(m_b)|^2 + \nonumber\\
&& |C_{Q_2}(m_b) - C^\prime_{Q_2}(m_b) + 2\hat{m}( C_{10}(m_b) -
C^\prime_{10}(m_b) )|^2]
\end{eqnarray}
where $\hat{m} = m_\mu/m_{B_s}$. With large $C^{(\prime)}_{Q_{1,2}}$,
Br($B_s \rightarrow \mu^+ \mu^-$) can be enhanced by several order of
magnitude.

\section{Squark mass matrices}\label{sect:squark}
The $6 \times 6$ squark mass-squared matrices in the super-CKM basis
have the structure
\begin{equation}
\label{squark:mass}
\M^2_{\tilde{q}} = \left(\begin{array}{cc}
\M^2_{\tilde{q}_{LL}} & \M^2_{\tilde{q}_{LR}}\\
\M^{2\dagger}_{\tilde{q}_{LR}} & \M^2_{\tilde{q}_{RR}}
\end{array}\right)
\end{equation}
where $\tilde{q} = \tilde{u} (\tilde{d})$ represent the up (down)-type squark.
Note that differently from $\M^2_{\tilde{q}_{LL,RR}}$,
$\M^2_{\tilde{q}_{LR}}$ is not hermitian.
In general, $\M^2_{\tilde{q}_{LL,RR,LR}}$
are off-diagonal. The $3\times 3$ submatrices are given by
\begin{eqnarray}
(\M^2_{\tilde{q}_{LL}})_{ij} &=& (m^2_{\tilde{q}_L})_{ij}
+ \delta_{ij}( {m^2_q}_i + \cos 2\beta m^2_Z (I^3_q - Q_q \sin^2 \theta_W) )
\nonumber\\
(\M^2_{\tilde{q}_{RR}})_{ij} &=& (m^2_{\tilde{q}_R})_{ij}
+ \delta_{ij}( {m^2_q}_i + Q_q \cos 2\beta m^2_Z \sin^2 \theta_W )
\nonumber\\
(\M^2_{u_{LR}})_{ij} &=& v_U {A_u}_{ij}
- \delta_{ij} {m_u}_i\mu\cot\beta \nonumber\\
(\M^2_{d_{LR}})_{ij} &=& v_D {A_d}_{ij}
- \delta_{ij} {m_d}_i\mu\tan\beta
\end{eqnarray}
where ${m_q}_i$ with $q=d,u$ is the quark mass of generation $i$,
$I^3_q$ and $Q_q$ are the third component of weak isospin and electric charge
of quark $q$ respectively, $\mu$ is the Higgs superfield mixing parameter,
${A_u}_{ij}$, ${A_d}_{ij}$ are trilinear higgs-squark-squark coupling.

Because of SU(2) gauge invariance, the squark mass matrix $m^2_{\tilde{u}_L}$ is
intimately connected to $m^2_{\tilde{d}_L}$ via
\begin{equation}
\label{su2}
m^2_{\tilde{u}_L} = V^\dagger_{\rm CKM} m^2_{\tilde{d}_L} V_{\rm CKM}
\end{equation}
Then, from $m^2_{\tilde{d}_L}$, we can get $m^2_{\tilde{u}_L}$,  and vice versa.

Furthermore, we assume for the sake of simplicity that there are
no new CP-violating phases, besides the single CKM phase.
Thus, we in general have twenty seven new flavor violation parameters
totally from squark mass matrice.
However, only five of them are involved in the transition b to s
in our analysis (see below).

Because we concentrate on the $b-s$ transition, only left-left, right-right
and left-right $2-3$ mixing terms are directly dependent.
In order to simplify the analysis we shall keep only
these $2-3$ mixing terms non-zero and
set all the $1-2$ and $1-3$ mixing terms to $0$.
We also keep the third generation $3-3$ left-right mixing term non-zero.
The first generation $1-1$ and second generation $2-2$ left-right
mixing terms are set to $0$ for simplicity.
We parametrize the non-vanishing $2-3$ off-diagonal term as
\begin{eqnarray}
(m^2_{\tilde{d}_L})_{23} &=& \delta^{dLL}_{23}
\sqrt{ (\M^2_{\tilde{d}_{LL}})_{22} (\M^2_{\tilde{d}_{LL}})_{33} },\hspace{8mm}
(\M^2_{d_{LR}})_{23} = \delta^{dLR}_{23}
\sqrt{ (\M^2_{\tilde{d}_{LL}})_{22} (\M^2_{\tilde{d}_{LL}})_{33} } \nonumber\\
(\M^2_{d_{RL}})_{23} &=& \delta^{dRL}_{23}
\sqrt{ (\M^2_{\tilde{d}_{LL}})_{22} (\M^2_{\tilde{d}_{LL}})_{33} },\hspace{8mm}
(m^2_{\tilde{d}_R})_{23} = \delta^{dRR}_{23}
\sqrt{ (\M^2_{\tilde{d}_{RR}})_{22} (\M^2_{\tilde{d}_{RR}})_{33} } \nonumber\\
(\M^2_{u_{LR}})_{23} &=& \delta^{uLR}_{23}
\sqrt{ (\M^2_{\tilde{u}_{LL}})_{22} (\M^2_{\tilde{u}_{LL}})_{33} }
\end{eqnarray}

In the super-CKM basis the fields $\tilde{q}_{Li}$, $\tilde{q}_{Ri}$
($i=1,2,3$) are related to the mass eigenstates $\tilde{q}_a$ ($a=1,...,6$) by
\begin{equation}
\tilde{q}_{L,R}= \Gamma^{\dagger}_{q L,R} \tilde{q},
\end{equation}
Where the matrix $\Gamma_{q L,R}^{\dagger}$ ($q=U,D$) is a
$3\times 6$ mixing matrix.

In order to simplify the discussion further,
we assume all the diagonal elements of
squark mass matrices are equal to a commen SUSY scale at electroweak scale,
$m_{\tilde{q}} = m_{\tilde{u}_{Lii}} = m_{\tilde{u}_{Rii}}
= m_{\tilde{d}_{Lii}} = m_{\tilde{d}_{Rii}}$ ($i=1,2,3$).
Then we have three flavor conserved parameter $m_{\tilde {q}}$,
$\delta^{uLR}_{33}$, $\delta^{dLR}_{33}$ and
five flavor violation parameters involved
in the $b$ to $s$ transition, $\delta^{uLR}_{23}$,
$\delta^{dLL}_{23}$, $\delta^{dLR}_{23}$, $\delta^{dRR}_{23}$,
$\delta^{dRL}_{23}$, from the squark mass matrices.
In addition to these parameters,
we have also the following free parameters: gaugino masses $M_i$ (i=1,2,3),
CP-odd Higgs boson mass $M_{A^0}$, $\mu$, and $\tan\beta$.

\section{Numerical analysis of Wilson coefficients $C^{(\prime)}_{10}$
and $C^{(\prime)}_{Q_{1,2}}$}\label{sect:numerical}
As analyzed in section \ref{sect:hamiltonian}, $C^{(\prime)}_{10}$
from the $Z$ penguin and $C^{(\prime)}_{Q_{1,2}}$ from
neutral Higgs boson penguins can be significantly enhanced.
In this section, we present our numerical analysis of Wilson coefficients
$C^{(\prime)}_{10}$ and $C^{(\prime)}_{Q_{1,2}}$ versus
flavor violating parameters under experimental bounds,
particularly Br($B\rightarrow X_s \gamma$),
in the case of switching on only the gluino (or neutralino) and SM contributions
in the first three subsections (fourth section),
and then switching on all the contributions
from the W boson, charged Higgs, chargino, neutralino,
and gluino in the last subsection.
We show there are two cases, the heavy gluino,
and fine-tuning (i. e., to finely tune $\delta^{dLL(RR)}_{23}$ and
$\delta^{dLR(RL)}_{23}$ terms in $C^{(\prime)}_7$ makes the constraint
from $b\rightarrow s\gamma$ satisfied) to enhance $C^{(\prime)}_{10}$.
In general $C_9^{(\prime)}(m_b)$ in MSSM is enhanced by at most five persent
compared to SM. We can see the reason why it can not have a large
enhancement from the expressions of $C_9^{(\prime)}$
in the Appendix. The gamma penguin and box contributions to $C_9^{(\prime)}$
are suppressed by a factor of
$m_w^2/m^2_{\tilde{\chi}^\pm,\tilde{\chi}^0,\tilde{g}}$. And the Z- penguin
contributions to $C_9^{(\prime)}$ are suppressed by $(-1 + 4 s_w^2)$.
So we shall not discuss its dependence on new flavor violation parameters
in most of part of the numerical analysis hearafter.
In our numerical analysis we use the expressions of Wilson coefficients
obtained by the VM method, i. e., those given in Appendix.
In order to simplify our notation and express
new physics effects  we introduce the following quantities,
\begin{eqnarray}
R_i = \frac{C^{\rm mssm}_i(m_w) - C^{\rm sm}_i(m_w)}{C^{\rm sm}_i(m_w)},
\hspace{5mm}
R^\prime_i = \frac{C^{\prime\rm mssm}_i(m_w)}{C^{\rm sm}_i(m_w)}
\end{eqnarray}
where $C_i^{(\prime)\rm mssm}$ is the Wilson coefficient of the operator
$O_i^{(\prime)}$ in MSSM and $C_i^{\rm sm}$ in SM,  and $i=7,8,9,10$.

The experimental measurements of mass differences in
$\bar{B}^0_s-B^0_s$ system give the following bound~\cite{sto}
\begin{equation}
\Delta M_s \ge 14.4 {\rm ps}^{-1}\,.
\end{equation}
We consider the constraints of $\Delta M_s$ on $\delta$'s as in
ref.~\cite{chang01}\footnote{The analysis in ref.~\cite{chang01}
is based on the old experimental bound, $\Delta M_s \ge 15.0 {\rm
ps}^{-1}$~\cite{lepbosc}. An update analysis in MSSM is in
progress~\cite{massd}.}. Current experimental measurements of
inclusive decay $B \rightarrow X_s \gamma$ at
ALEPH~\cite{ALEPHbsgamma}, CLEO~\cite{CLEObsgamma} and
BELLE~\cite{BELLEbsgamma} produce the world average value
\begin{equation}
{\rm Br}(B \rightarrow X_s \gamma)_{\rm exp} = (3.23 \pm 0.41) \times 10^{-4}\,.
\end{equation}
We use $2\sigma$ Br($B\rightarrow X_s \gamma$) bound in our numerical analysis.

\subsection{Heavy gluino contributions to $C^{(\prime)}_{10}$}
\begin{figure}
\begin{tabular}{cc}
\epsfxsize=8cm
\epsfysize=8cm
\epsffile{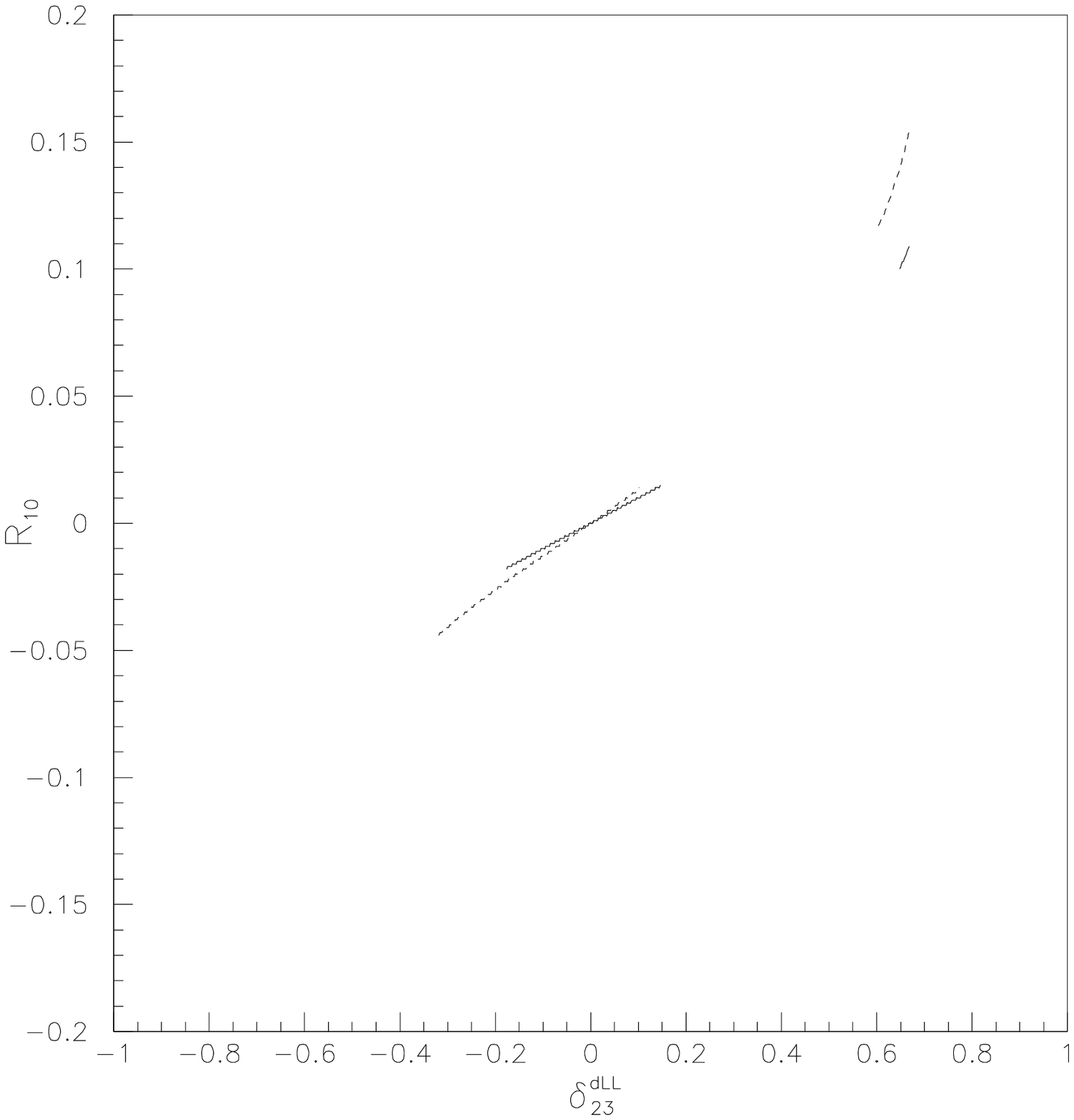} &
\epsfxsize=8cm
\epsfysize=8cm
\epsffile{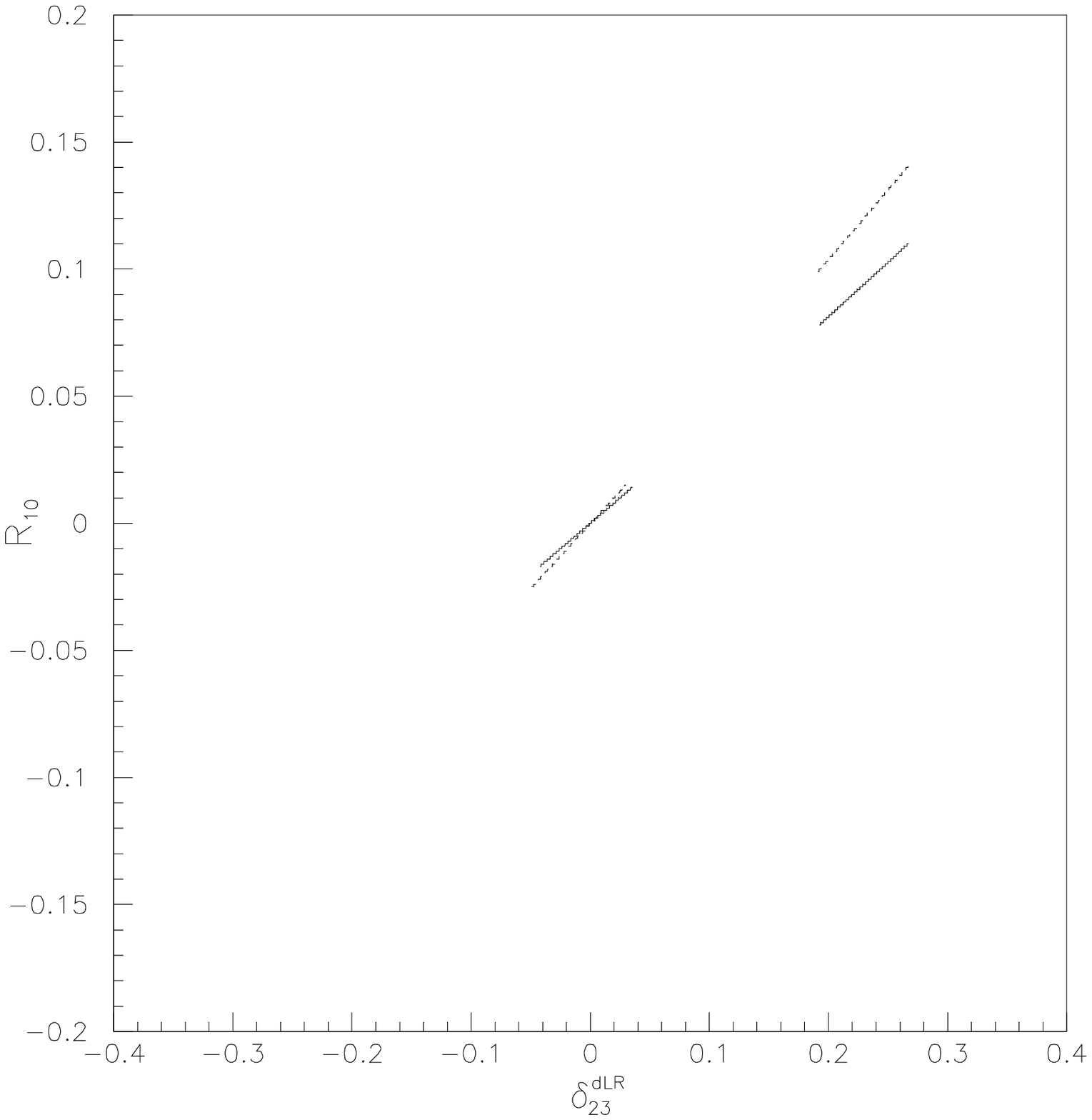}\\
a & b\\
\epsfxsize=8cm
\epsfysize=8cm
\epsffile{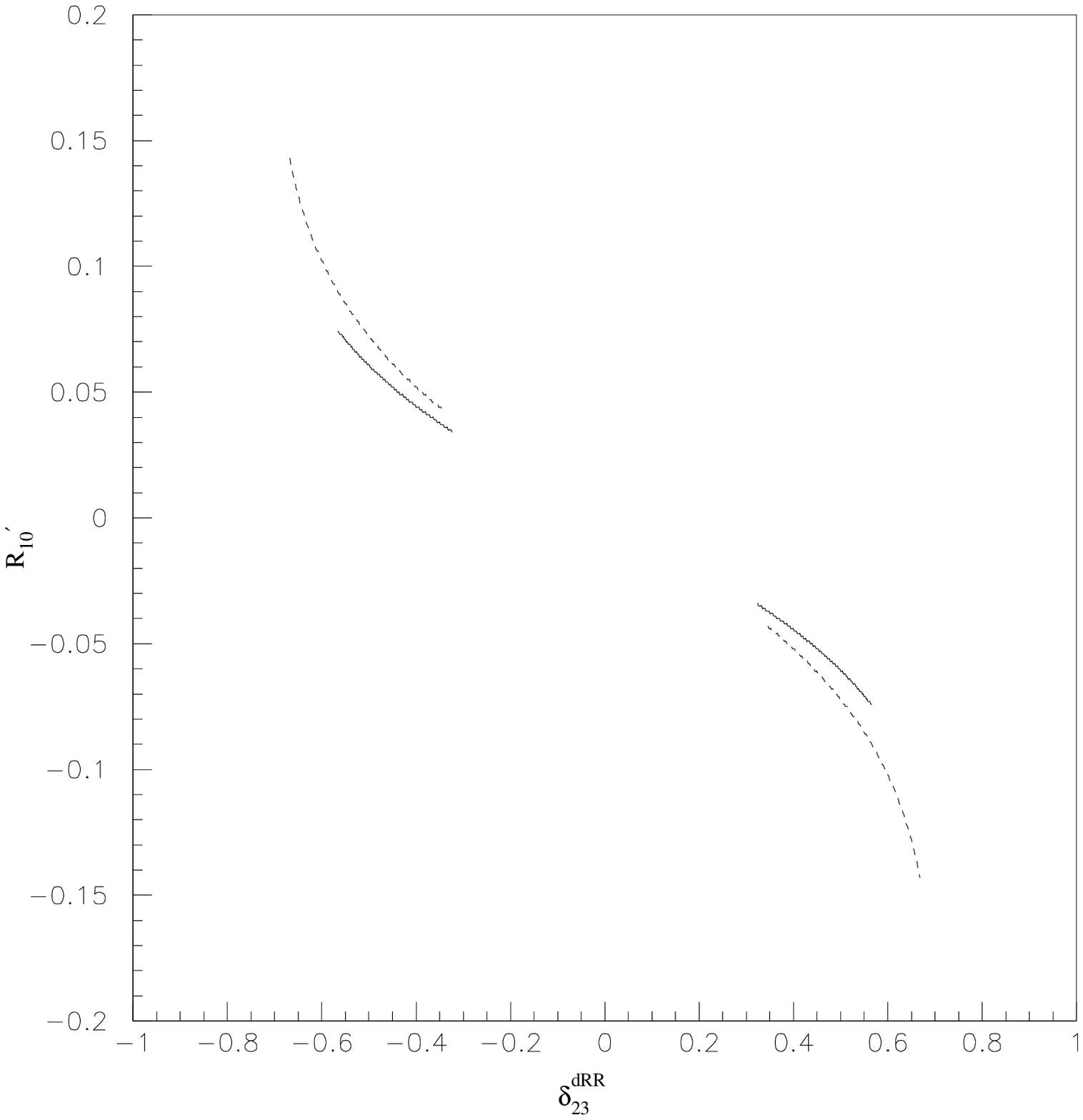} &
\epsfxsize=8cm
\epsfysize=8cm
\epsffile{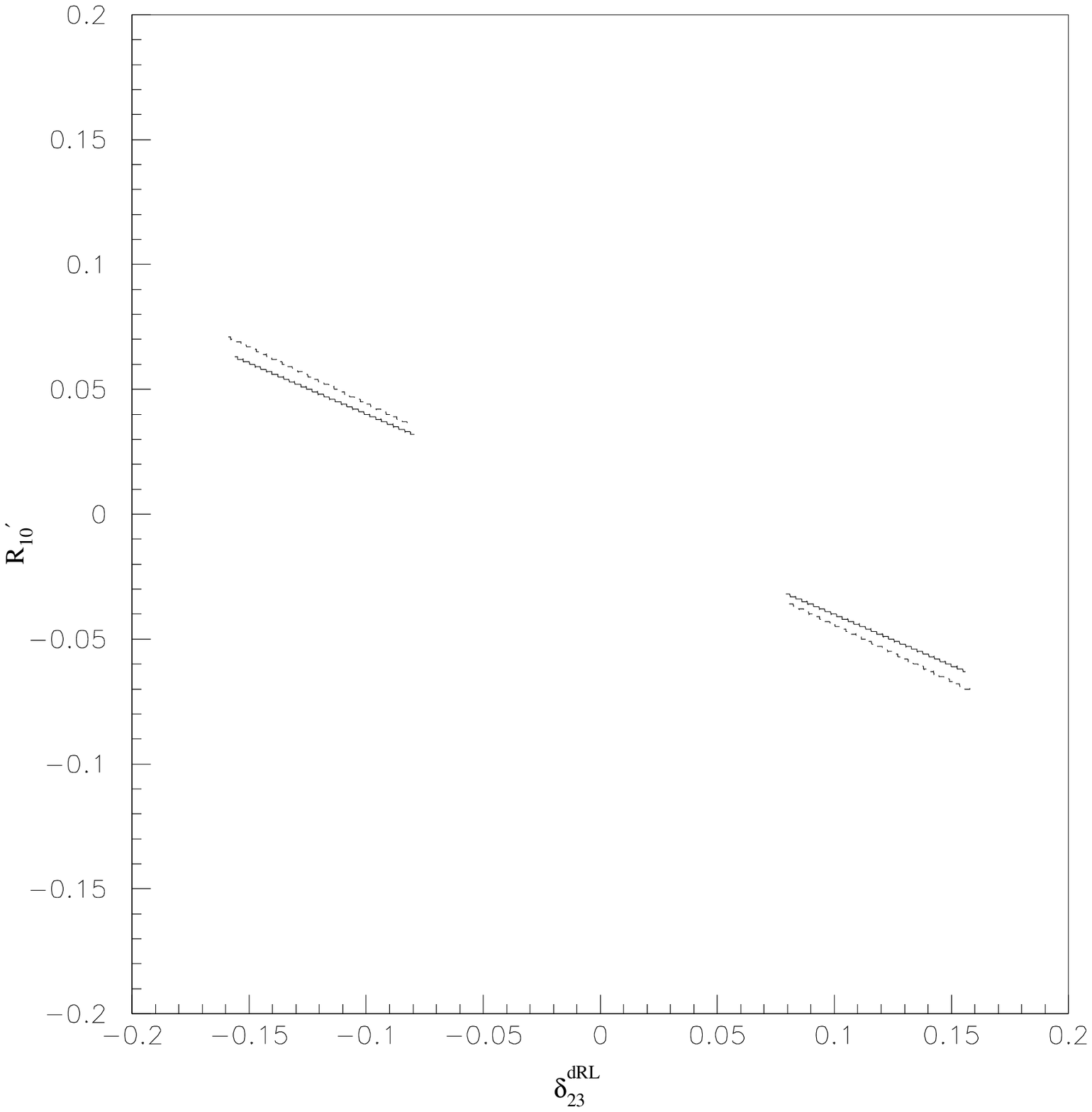}\\
c & d
\end{tabular}
\caption{Dependence of $R^{(\prime)}_{10}$ on $\delta^{dLL,dLR(dRR,dRL)}_{23}$.
Gluino and SM contributions are denoted by solid curves, and dot curves
denote all the contribution.
The other parameters are $M_{\tilde{q}} = 800$GeV,
$M_3 = 3000$GeV, $M_2 = 1200$GeV, $M_1 = 100$GeV,
 $\mu = 3200$GeV and $\tan\beta = 50$.
}
\label{gheavy}
\end{figure}

As we have noticed in section \ref{sect:hamiltonian},
though $C^{(\prime)}_7$ has $\frac{m_{\tilde{g}}}{m_b}$ enhancement,
it still suffers from $\frac{m^2_w}{m^2_{\tilde{g}}}$ suppression.
Then $C^{(\prime)}_7$ is suppressed by $\frac{1}{m_{\tilde{g}}}$,
as presented in Eq.(\ref{c7}).
Nevertheless the $Z$ penguin contribution to $C^{(\prime)}_{10}$
is non-decoupled when $m_{\tilde{g}}$ is large. Therefore,
we expect that when the gluino is heavy,
the $b\rightarrow s\gamma$ constraint can be easily satisfied and
a large enhancement of $C^{(\prime)}_{10}$ can occur.
Including, in addition to the SM contribution, only the gluino contribution,
we show $\delta^{dLL}_{23}$, $\delta^{dLR}_{23}$, $\delta^{dRR}_{23}$,
and $\delta^{dRL}_{23}$  dependences of $C_{10}^{(\prime)}$
respectively, i. e., in each case only one non-zero off-diagonal parameter
enters, in Fig.~\ref{gheavy}, where
$M_{\tilde{q}} = M_{\tilde{l}} = 800$GeV, $M_1 = 100$GeV,
$M_2 = 1200$GeV, $M_3 = 3000$GeV, $\mu = 3200$GeV and $\tan\beta = 50$.
The choice of large $\mu$ and $\tan\beta$ is to ensure large left-handed
and right-handed sbottom mixing, $\delta_{33}^{dLR}= -0.75$.

In Fig.~\ref{gheavy}.$a$, we give $R_{10}$ as
a function of $\delta^{dLL}_{23}$. When $\delta^{dLL}_{23}$ is near $0.6$,
$R_{10}$ is the order of $10\%$, where the sign of $C_7(m_b)$ is flipped,
and the lightest sbottom mass is near the low bound obtained in  experiments.
We show $R_{10}$ as a function of $\delta^{dLR}_{23}$ in Fig.~\ref{gheavy}.$b$.
$R_{10}$ can also reach the order of $10\%$.

When we switch on only $\delta^{dRR}_{23}$ or $\delta^{dRL}_{23}$,
we assume some mechanism to render $C_7(m_b)$ to be $0$, e.g.,
some non-zero $\delta^{dLL}_{23}$ or $\delta^{dLR}_{23}$
contribution to $C_7(m_w)$ nearly cancells the SM contribution.
We present $R^\prime_{10}$ as a function of
$\delta^{dRR}_{23}$ in Fig.~\ref{gheavy}.$c$. When $\delta^{dRR}_{23}$
is near $-0.6$, $R^\prime_{10}$ can be as large as $7\%$.
$R^\prime_{10}$ as a function of $\delta^{dRL}_{23}$
are shown in Fig.~\ref{gheavy}.$d$. When $\delta^{dRL}_{23}$
is near $-0.15$, $R^\prime_{10}$ can be as large as $7\%$.

\subsection{Fine-tuning of $\delta^{dLL(RR)}_{23}$ and $\delta^{dLR(RL)}_{23}$
terms in $C^{(\prime)}_7$ and gluino contributions to $C^{(\prime)}_{10}$}
It is obvious from Eq.(\ref{c7}) that $C^{(\prime)}_7$ can
be finely tuned to zero with the large cancellation between
$\delta^{dLL(RR)}_{23}$ and $\delta^{dLR(RL)}_{23}$ terms.
However, in this case the $\delta^{dLL(RR)}_{23}$ and
$\delta^{dLR(RL)}_{23}$ contributions to $C^{(\prime)}_{10}$ can be large.

\begin{figure}
\epsfxsize=18cm
\epsfysize=18cm
\epsffile{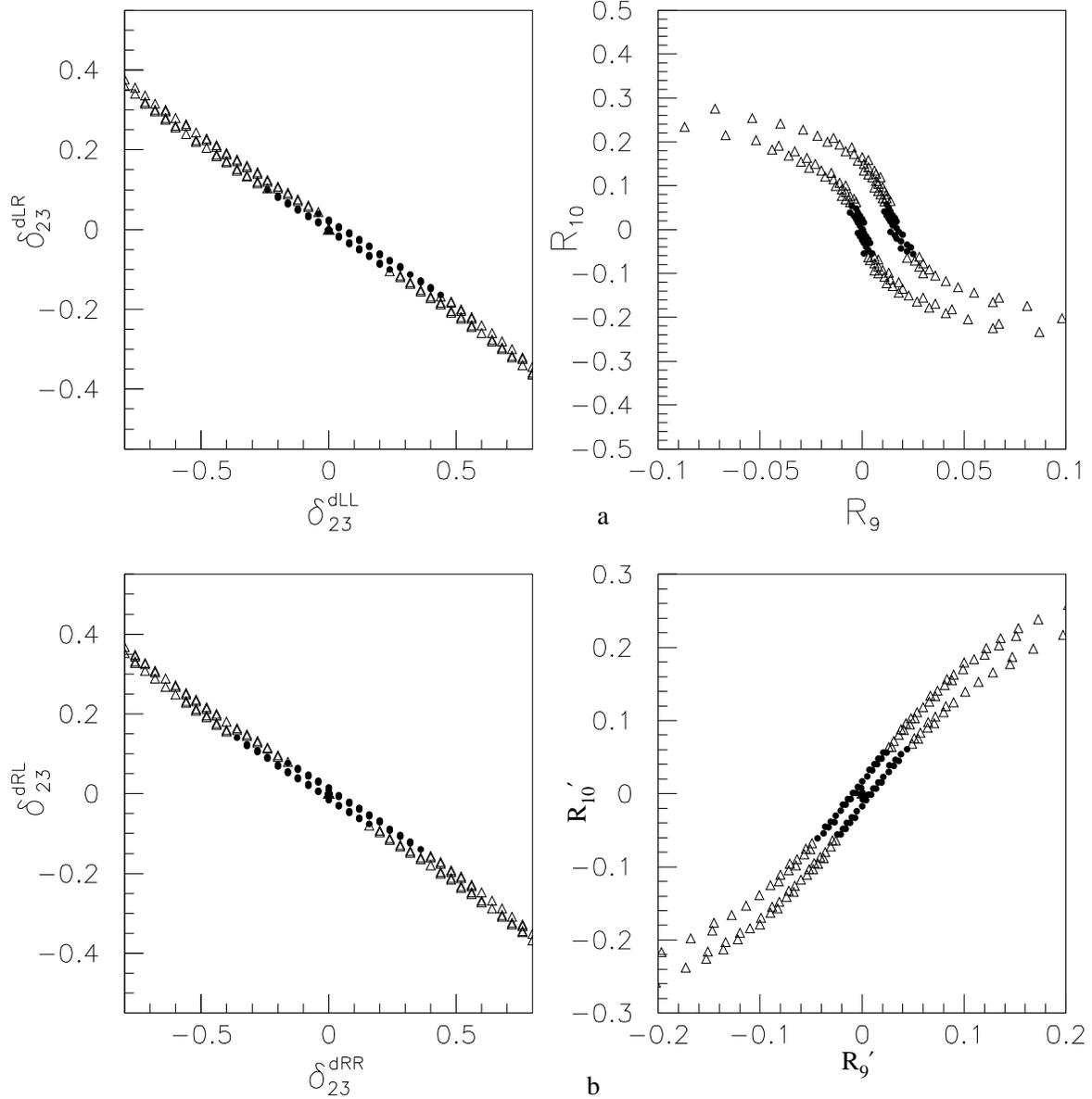}
\caption{Correlations between $\delta^{dLL(dRR)}_{23}$ and $\delta^{dLR(dRL)}_{23}$ and between  $R^{(\prime)}_9$ and $R^{(\prime)}_{10}$
switching on only gluino and SM contributions to $R^{(\prime)}_9$ and $R^{(\prime)}_{10}$
in the fine-turning case.
The other parameters are $M_{\tilde{q}} = 500$GeV,
$M_3 = 500$GeV, $\mu = 1200$GeV, and $\tan\beta = 50$.
Triangle points are ruled out by $B \rightarrow X_s g$ and hadronic charmless
$B$ decays which require $|R_8| \le 10$.
}
\label{gfinetune}
\end{figure}

We show the correlations of $\delta^{dLL}_{23}$ {\it vs}
$\delta^{dLR}_{23}$ and $R_{9}$ {\it vs} $R_{10}$
in Fig.~\ref{gfinetune}.$a$, where fine-tuning of $\delta^{dLL}_{23}$
and $\delta^{dLR}_{23}$ terms in $C_7$ have been carried out.
The other parameters are $M_{\tilde{q}} = 500$GeV,
$M_3 = 500$GeV, $\mu = 1200$GeV, and $\tan\beta = 50$.
Large $\mu$ and $\tan\beta$ is to ensure large sbottom mixing
$\delta_{33}^{dLR}= -0.72$.
We find that the largest $R_{10}$
is $40\%$. However, when we impose the constraints from
$B \rightarrow X_s g$ and hadronic charmless B decays which require
$|R_8|$ cannot be larger than
$10$~\cite{greub00,ali01}, $R_{10}$ can only be as large as $8\%$.

In order to analyze $\delta^{dRR(RL)}_{23}$, we set $C_7(m_b)$
to $0$ and keep $C_9(m_b)$ and $C_{10}(m_b)$ to be their
corresponding SM values. We make the fine-tuning of $\delta^{dRR}_{23}$
and $\delta^{dRL}_{23}$ in $C^{\prime}_7$ so that
the constraint from $b\rightarrow s
\gamma$ is satisfied. With $M_{\tilde{q}} = 500$GeV,
$M_3 = 500$GeV, $\mu = 1200$GeV, $\tan\beta = 50$,
and sbottom mixing $\delta_{33}^{dLR}= -0.72$, the gluino
contributions to $R^\prime_9$ and
$R^\prime_{10}$ are shown in Fig.~\ref{gfinetune}.$b$.
$R^{\prime}_{10}$ can reach $40\%$ and $8\%$ without and
with the constraints from
$B \rightarrow X_s g$ and hadronic charmless B decays respectively.

\subsection{Gluino contributions  to  $C^{(\prime)}_{Q_{1,2}}$}
As stressed in section \ref{sect:hamiltonian}, when the CP-odd Higgs
is not too heavy and $\tan\beta$ is large,
neutral Higgs contribute significantly via
$C^{(\prime)}_{Q_{1,2}}$ to the process (\ref{bs}).
As we have stated before, only $\delta^{dLL(RR)}_{23}$ have sizable
contribution to $C^{(\prime)}_{Q_{1,2}}$. The contribution from
$\delta^{dLR(RL)}_{23}$ can be neglected due to the cancellation between
contributions of the self energy type and penguin diagrams.
Because the loop functions decease when the mass of gluino increases,
gluino contributions  to $C^{(\prime)}_{Q_{1,2}}$ in the heave gluino case
are less significant than those in the not too heavy gluino case.
So in this subsection we consider only the case where gluino is not too heavy.

We show in Fig.~\ref{ghiggs}.$a$ the gluino contribution to $C_{Q_1}$ {\it vs}
$\delta^{dLL}_{23}$ with $M_{\tilde{q}} = 500$GeV, $M_{A^0} = 250$GeV,
$\mu = 800$GeV, $\tan\beta = 40$, and sbottom left-right mixing
$\delta^{dLR}_{33} = -0.38$.
The other parameters are $M_3 = 500$GeV and the SU(5) gaugino mass relation
at electroweak scale $M_Z$, $M_1 : M_2 : M_3 = 1 : 2 : 7$ for simplicity.
$C_{Q_1}(m_b)$ can be as large as $2.5$,
when $\delta^{dLL}_{23}$ is near $0.04$.
We can see that due to the constraint of Br($B \rightarrow X_s \gamma$),
$\delta^{dLL}_{23}$ is restricted to the order of $0.04$.

In order to analyze $C^\prime_{Q_{1,2}}$, we set $C_7(m_b)$ and $C_8(m_b)$
to $0$, and keep $C_9(m_b)$ and $C_{10}(m_b)$ unchanged from their SM values.
We show in Fig.~\ref{ghiggs}.$b$ the gluino contribution
to $C^\prime_{Q_1}$ {\it vs}
$\delta^{dRR}_{23}$ with $M_{\tilde{q}} = 500$GeV, $M_{A^0} = 250$GeV,
$M_3 = 500$GeV, $\mu = 800$GeV, $\tan\beta = 40$, and sbottom mixing
$\delta^{dLR}_{33} = -0.38$. In the figure the solid and dot curves
(which denote the gluino contribution and all contributions respectively)
are almost overlapping.
$C^\prime_{Q_1}(m_b)$ can be as large as $3.6$,
when $\delta^{dRR}_{23}$ is near $0.06$.

\subsection{Neutralino contributions to $C_{10}^{(\prime)}$
and $C^{(\prime)}_{Q_{1,2}}$}
\begin{figure}
\epsfxsize=18cm
\epsfysize=18cm
\epsffile{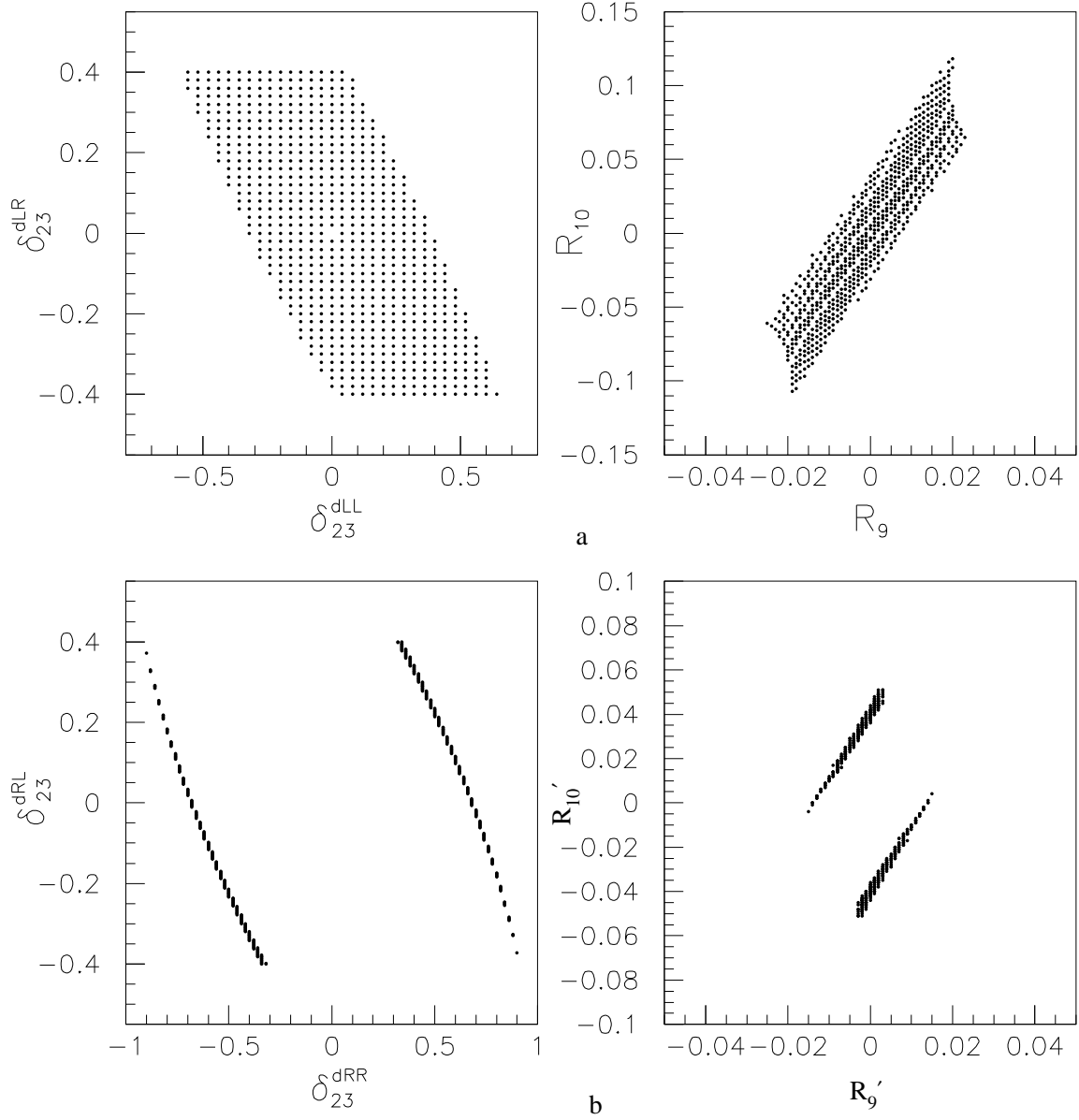}
\caption{Correlations between $\delta^{dLL(dRR)}_{23}$ and $\delta^{dLR(dRL)}_{23}$ and between  $R^{(\prime)}_9$ and $R^{(\prime)}_{10}$
switching on only neutralino and SM contributions to $R^{(\prime)}_9$ and $R^{(\prime)}_{10}$
in the fine-turning case.
The other parameters are $M_{\tilde{q}} = 500$GeV,
$M_1 = 80$GeV, $M_2 = 300$GeV, $\mu = 1200$GeV, and $\tan\beta = 50$.
}
\label{chifinetune}
\end{figure}

In the heavy gluino case, because we choose $M_1=100$GeV,
the lightest neutralino should be light so that it contributes to
$R_{10}^{(\prime)}$ greatly and constructively with the gluino contribution,
as can be seen from the dot curve in Fig.~\ref{gheavy}
which corresponds the case of
including all contributions. We now turn to the fine-tuning case.
In the case neutralinos can also has large effect
as long as the lightest neutralino is light enough.
We show in Fig.~\ref{chifinetune}.$a$ the correlation of
$\delta^{dLL}_{23}$ {\it vs} $\delta^{dLR}_{23}$
and $R_9$ {\it vs} $R_{10}$.
With $M_{\tilde{q}} = 500$GeV, $M_1 = 80$GeV,
$M_2 = 300$GeV, $\mu = 1200$GeV, $\tan\beta = 50$,
sbottom mixing $\delta^{dLR}_{33} = -0.72$,
$R_{10}$ can  reach $10\%$. And $R_9$ can reach only $3\%$, as expected.
We show the neutralino contributions to $R^{\prime}_9$ and
$R^\prime_{10}$ and the correlation
$\delta^{dRR}_{23}$ {\it vs} $\delta^{dRL}_{23}$
in Fig.~\ref{chifinetune}.$b$ with the same values of parameters
as those  in Fig.~\ref{chifinetune}.$a$.
$R^\prime_{10}$ can reach the order of $6\%$.
And $R_9^{\prime}$ can reach only $2\%$.

Neutralinos can also have large contributions to $C^{(\prime)}_{Q_{1,2}}$.
The neutralino contribution to $C_{Q_1}$ {\it vs} $\delta^{dLL}_{23}$,
with $M_{\tilde{q}} = 500$GeV, $M_{A^0} = 250$GeV, $M_1 = 100$GeV,
$M_2 = 300$GeV, $M_3 = 1000$GeV, $\mu = 800$GeV, $\tan\beta = 40$,
and sbottom mixing $\delta^{dLR}_{33} = -0.38$, is shown in Fig.~\ref{chihiggs}.
The $C_{Q_1}(m_b)$ can be as large as $1.5$,
when $\delta^{dLL}_{23}$ is near $-0.7$.
It is too hard for the non-zero $\delta^{dRR}_{23}$ to
generate sizable neutralino contributions to
$C^\prime_{Q_{1,2}}$, under the severe Br($B\rightarrow X_s \gamma$)
experimental constraint.

\subsection{Wilson coefficients $C^{(\prime)}_{10}$
and $C^{(\prime)}_{Q_{1,2}}$ in MSSM}
In this subsection we switch on all the contributions of charged gauge boson
$W^\pm$, charged Higgs $H^\pm$, chargino $\tilde{\chi}^\pm$,
neutralino $\tilde{\chi}^0$ and gluino $\tilde{g}$ and all the
$\delta$s, $\delta^{uLR}_{23}$, $\delta^{dLL}_{23}$,
$\delta^{dLR}_{23}$, $\delta^{dRR}_{23}$, and $\delta^{dRL}_{23}$,
under the experimental constraints.
Specifically, we are interested in calculating the maximum enhancements of
the Wilson coefficients $C^{(\prime)}_{10}$ and $C^{(\prime)}_{Q_{1,2}}$
that SUSY can provide, which is important to discriminate
the small $\tan\beta$ and large $\tan\beta$ scenarios by the measurement
of Br($B_s\rightarrow \mu^+\mu^-$).

\begin{figure}
\begin{tabular}{cc}
\epsfxsize=8cm
\epsfysize=8cm
\epsffile{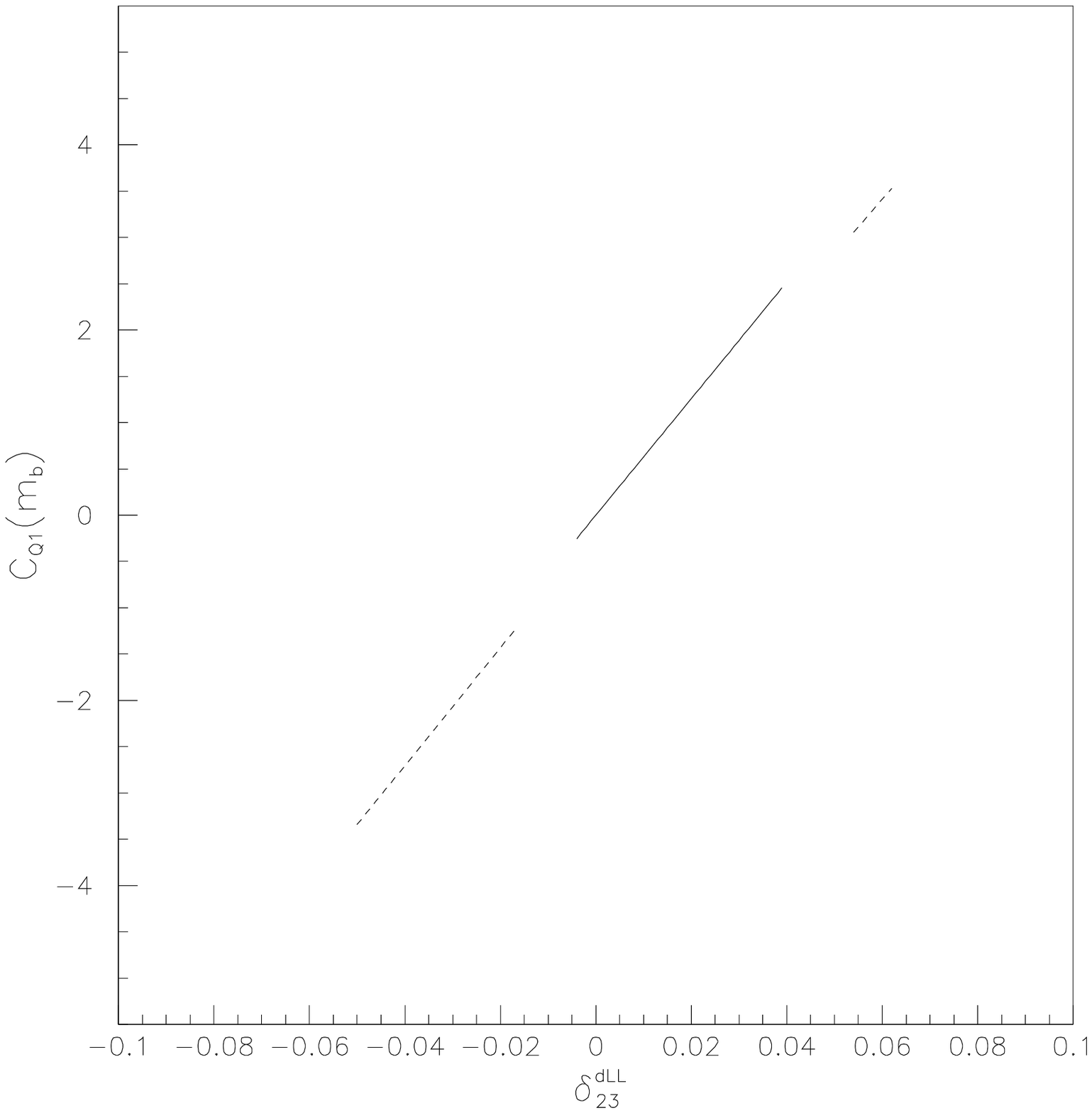} &
\epsfxsize=8cm
\epsfysize=8cm
\epsffile{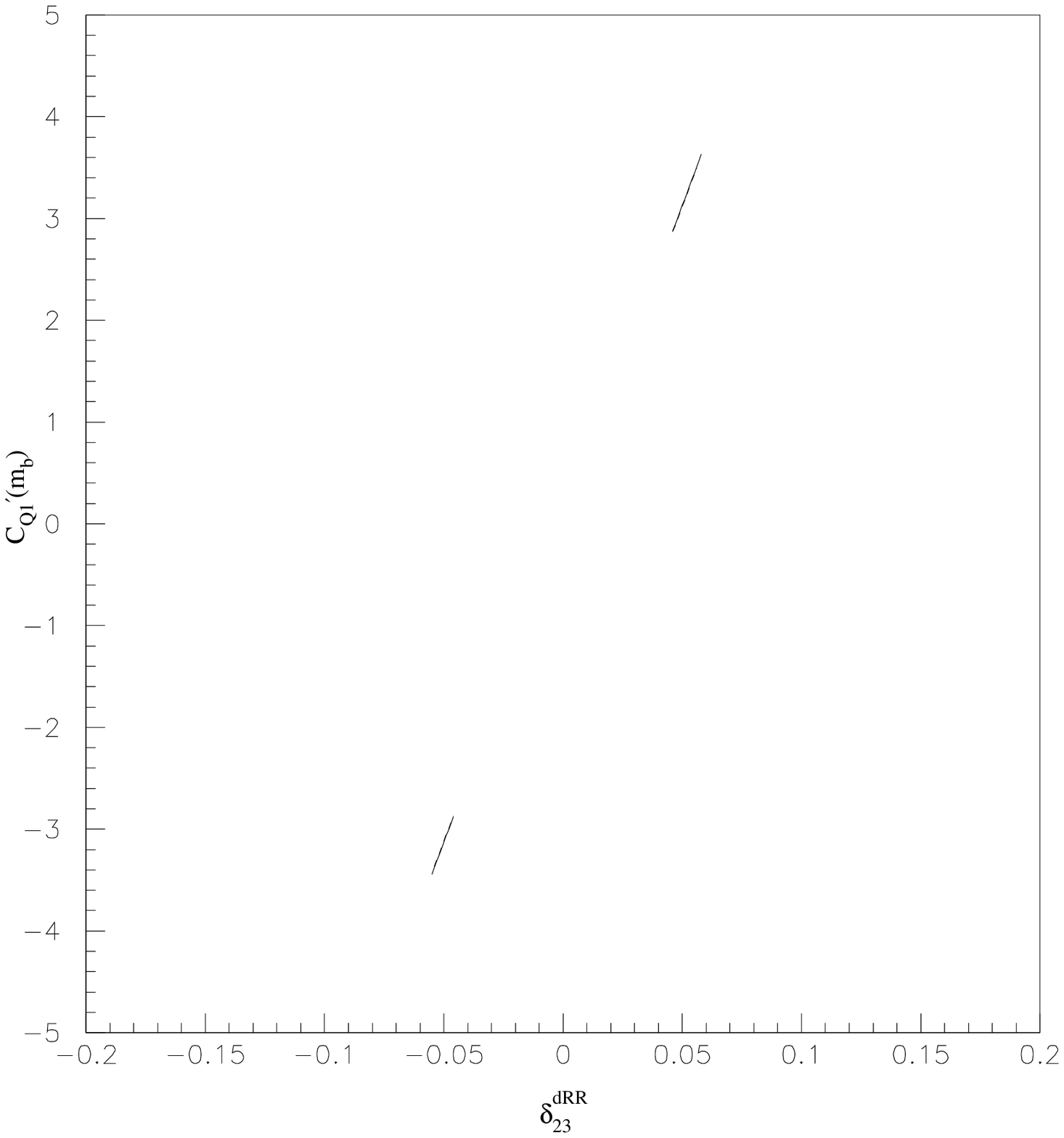} \\
a & b
\end{tabular}
\caption{Dependence of $C^{(\prime)}_{Q_1}(m_b)$ on $\delta^{dLL(dRR)}_{23}$.
Gluino and SM contributions are denoted by solid curves, and dot curves
denote all the contribution.
The parameters are $M_{\tilde{q}} = 500$GeV,
$M_{A^0} = 250$GeV, $\mu = 800$GeV and $\tan\beta = 40$ as well as
$M_3 = 500$GeV and the SU(5) gaugino mass relation
at the electroweak scale $M_Z$, $M_1 : M_2 : M_3 = 1 : 2 : 7$.
}
\label{ghiggs}
\end{figure}

\begin{figure}
\epsfxsize=8cm
\epsfysize=8cm
\centerline{\epsffile{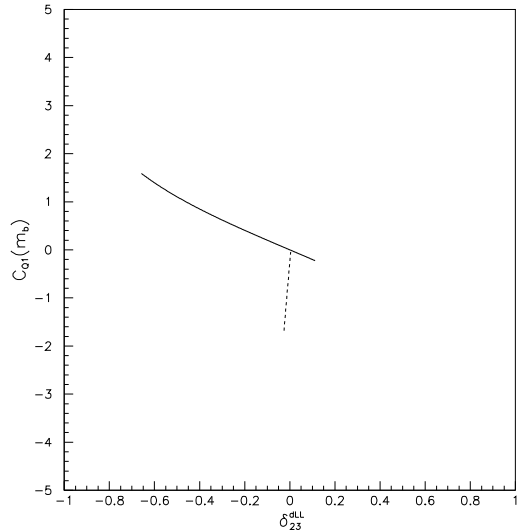}}
\caption{Dependence of $C_{Q_1}(m_b)$ on $\delta^{dLL}_{23}$.
The solid curve denotes neutralino and SM contributions, and the dot curve
denotes all the contributions.
The other parameters are $M_{\tilde{q}} = 500$GeV, $M_{A^0} = 250$GeV,
$M_1 = 100$GeV, $M_2 = 300$GeV, $M_3 = 1000$GeV, $\mu = 800$GeV
and $\tan\beta = 40$.
}
\label{chihiggs}
\end{figure}

First, we limit ourself to the case of only one non-zero
$\delta_{23}^{dAB}$ and including all the contributions
under the experimental constraints. The results are
presented in Figs.~\ref{gheavy}, \ref{ghiggs} and \ref{chihiggs}.
The dependences of $C^{(\prime)}_{10}$ on $\delta^{,}$s are shown
by dot curves in Fig.~\ref{gheavy}.
It is clear that the allowed ranges of $\delta^{dLL(RR)}_{23}$
and $R^{(\prime)}_{10}$ are enlarged and
$R^{(\prime)}_{10}$ can reach $15\%$.
The dependences of $C^{(\prime)}_{Q_{1,2}}$ on $\delta^{,}$s are
shown by dot curves in Fig.~\ref{ghiggs} and Fig.~\ref{chihiggs}
with different values of gaugino masses.
In the case of $M_3 = 500 $ GeV, because non-zero $\delta^{dLL}_{23}$
can induce scalar up-type quark $2-3$ left-left mixing,
the chargino can have large contributions and $C^{(\prime)}_{Q_{1,2}}$
can reach about $3.5$,
as shown in the Fig.~\ref{ghiggs}.$a$.
As for the dependence on $\delta^{dRR}_{23}$,
not like the non-zero $\delta^{dLL}_{23}$ case, both the chargino and
neutralino contributions are small, as shown in the Fig.~\ref{ghiggs}.$b$.
In the case of $M_3=1000$ GeV, both chargino and gluino contributions
are important, $C_{Q_{1,2}}$ can reach about $-2$,
and  the dependence on $\delta^{dLL}_{23}$ is quite different from that
in the case of switching on only the neutralino contribution,
as shown in Fig.~\ref{chihiggs}.

\begin{figure}
\epsfxsize=18cm
\epsfysize=18cm
\epsffile{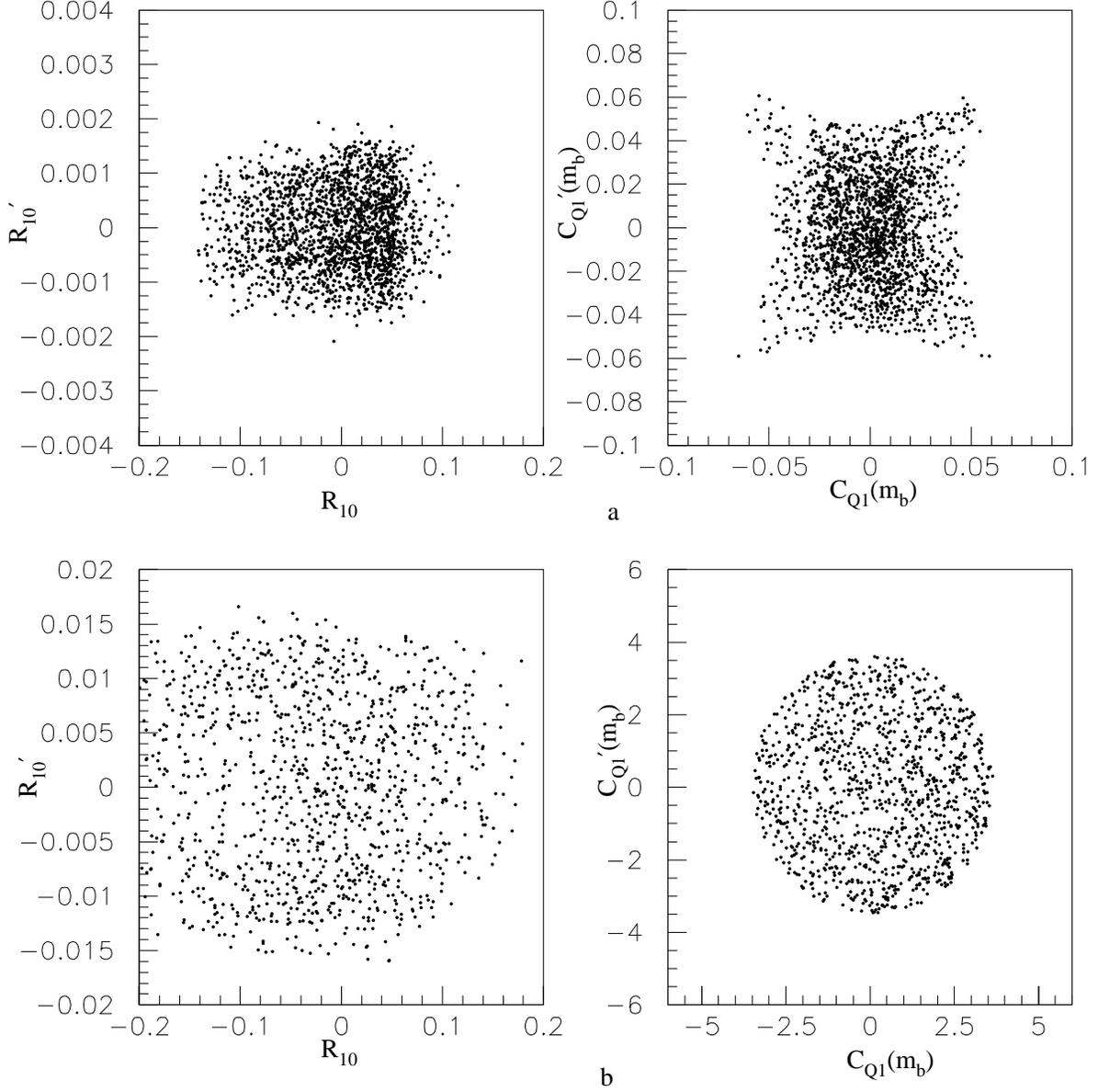}
\caption{Correlations between $R_{10}$ {\it vs} $R^\prime_{10}$ and between
$C_{Q_1}$ {\it vs} $C^\prime_{Q_1}$ switching on all the contributions and all $\delta^{qAB}_{23}$'s.
(a)  and (b) for $\tan\beta$=4 and 50 respectively.
The other parameters are $M_{\tilde{q}} = 500$GeV, $\mu = 500$GeV, $M_3 = 1000$GeV and the SU(5) gaugino mass relation
at the electroweak scale $M_Z$, $M_1 : M_2 : M_3 = 1 : 2 : 7$,
}
\label{overall}
\end{figure}

Next, we switch on all the contributions and all the
$\delta$s, $\delta^{uLR}_{23}$, $\delta^{dLL}_{23}$,
$\delta^{dLR}_{23}$, $\delta^{dRR}_{23}$, and $\delta^{dRL}_{23}$,
under the experimental constraints.
We perform a Monte Carlo scan of the parameter
space with the ranges, $|\delta^{uLR}_{23}| \le 1$,
$|\delta^{dLL(RR)}_{23}| \le 1$, $|\delta^{dLR(RL)}_{23}| \le 0.02$.
There are three gaugino masses $M_{1,2,3}$ which are free parameters in MSSM.
For simplicity, we employ the
SU(5) gaugino mass relation at the electroweak scale $M_Z$,
$M_1 : M_2 : M_3 = 1 : 2 : 7$ and take $M_3=1000$GeV in the scan.
We fix $M_{\tilde{q}}$, $\mu$, $A_{(u,d)33}$
with $M_{\tilde{q}} = 500$GeV, $\mu = 500$GeV, $A_{u33} = 250$GeV, and
$A_{d33} = 0$ and consider two cases,
small $\tan\beta = 4$ and large $\tan\beta = 50$.
The results of the scan are presented in
correlations, $R_{10}$ {\it vs} $R^\prime_{10}$ and
$C_{Q_1}$ {\it vs} $C^\prime_{Q_1}$, in Fig.~\ref{overall}.
In the small $\tan\beta = 4$ case, $R_{10}$ can reach $12\%$,
$R^\prime_{10}$ is nearly zero, and
$|C_{Q_1}(m_b)|$ and $|C^\prime_{Q_1}(m_b)|$ are smaller than $0.1$
in some parameter region due to the smallness of $\tan\beta$.
In the large $\tan\beta$ case,
$R_{10}$ can reach $20\%$, $R^\prime_{10}$ can reach $1.5\%$,
and $C_{Q_1}(m_b)$ and $C^\prime_{Q_1}(m_b)$ can be as large as $3.6$
in some parameter region.

Only six Wilson coefficients, $C^{(\prime)}_{10}$,
$C^{(\prime)}_{Q_{1,2}}$, affect Br($B_s \rightarrow \mu^+ \mu^-$).
Because $C^{(\prime)}_{Q_{1,2}}$ cannot be large
in the small $\tan\beta$ case, the main contribution to
Br($B_s \rightarrow \mu^+ \mu^-$) comes from $C^{(\prime)}_{10}$
which can have at most an enhancement of
order of $20\%$ compared with the Standard Model $C_{10}$.
But in the large $\tan\beta$ case, if the CP-odd Higgs boson is not too heavy,
$C^{(\prime)}_{Q_{1,2}}$ can be as large as order of one, which strongly
enhances Br($B_s \rightarrow \mu^+ \mu^-$) by a factor of $10^1-10^3$
depending on the values of parameters,
as shown in next section.

\section{Numerical results for $B_s \rightarrow \mu^+\mu^-$ and
$B \rightarrow X_s \mu^+ \mu^-$}\label{sect:bsmu}

In the numerical calculations, the following input parameters have
been used:

 $\alpha_s(m_z)=0.118$, $\alpha_s(m_b)=0.215$,
$\alpha_{\rm em}(m_z)=\frac{1}{128}$, $\alpha_{\rm em}(m_b)=\frac{1}{132}$,
$\lambda_t=- 0.038$,\\ $V_{\rm cb}=0.04$,
Br($B \rightarrow X_c e \bar{\nu}$) = 0.11, $m_\mu=0.1057$GeV,
$m_b(m_z)=2.9$GeV, $m_b(m_b)=4.2$GeV,\\ $m_t^{\rm pole}=175$GeV,
$m_c^{\rm pole}/m_b^{\rm pole}=0.29$, $m_{B_s}=5.370$GeV,
$f_{B_s}=0.22$GeV, $\tau_{B_s}=1.46$ps.

As can be seen from eg. (\ref{slbr}), the branching ratio of $B_s
\rightarrow X_s \mu^+\mu^-$ depends on the sign of $C_7(m_b)$ and,
as mentioned before, in MSSM there exist some parameter regions,
where supersymmetric contributions to $C_7$ can make its sign
opposite to $C^{\rm SM}_7(m_b)$. Therefore, under the experimental
constraint from the branching ratio of $B\rightarrow X_s \gamma$,
two separate regions for the correlation between the branching
ratios of the $B\rightarrow X_s \gamma$ and $B\rightarrow X_s
\mu^+\mu^-$ are allowed. One corresponds to the case in which the
sign of $C_7(m_b)$ is the same as that in SM, and the other
corresponds to the the case when the sign of $C_7(m_b)$ is
opposite to that in SM, which is similar to the results given in
ref.~\cite{goto97}\footnote{In ref.~\cite{goto97} the neutral
Higgs boson contributions to the branching ratio of $B_s
\rightarrow \mu^+\mu^-$ in the large $\tan\beta$ case was missed
since they set the mass of a muon equal to zero.}. Our numerical
results verify the analysis.

As an illustration, in Tab.~\ref{tab:bsmu} we present numerical
results of Br($B_s \rightarrow \mu^+\mu^-$) and Br($B \rightarrow
X_s \mu^+ \mu^-$) for the two set of parameter values which are in
the region of the parameter space where the sign of $C_7(m_b)$ is
the same as that in SM. The other parameters which are not given
in the table are $M_3 = 1000$GeV and the SU(5) gaugino mass
relation at the electroweak scale $M_Z$, $M_1 : M_2 : M_3 = 1 : 2
: 7$, $M_{\tilde{q}} = 500$GeV, and $\mu = 500$GeV.

\begin{table}[!htb]
\caption{Wilson coefficients $C^{(\prime)}_{7,8,9,10,Q_{1,2}}$
and Branching ratios of $B\rightarrow X_s \mu^+ \mu^-$
and $B_s \rightarrow \mu^+ \mu^-$.
}
\label{tab:bsmu}
\begin{tabular}{rrrrrr}
case & $\delta^{uLR}_{23}$/$\delta^{dLL}_{23}$/$\delta^{dLR}_{23}$ &
$\delta^{dRR}_{23}$/$\delta^{dRL}_{23}$  & $C_7$/$C^\prime_7$ &
$C_8$/$C^\prime_8$ & $C_9$/$C^\prime_9$ \\
$\tan\beta$ & $C_{10}$/$C^\prime_{10}$ & $C_{Q_1}$/$C^\prime_{Q_1}$ &
$C_{Q_2}$/$C^\prime_{Q_2}$ & Br($B\rightarrow X_s \mu^+ \mu^-$) &
Br($B_s \rightarrow \mu^+ \mu^-$)\\
\hline
A & $0.79$/$-0.046$/$0.0039$ & $0.036$/$-0.0095$ &
$0.34$/$0.10$ & $0.15$/$0.06$ & $-4.4$/$-0.003$ \\
$4$ & $5.2$/$0.002$ & $-0.0011$/$0.0013$ &
$0.0008$/$0.0014$ & $4.28\times 10^{-6}$ & $4.29\times 10^{-9}$\\
\hline
B & $0.68$/$-0.044$/$0.018$ & $-0.037$/$-0.0065$ &
$0.40$/$0.16$ & $0.14$/$0.076$ &
$-4.39$/$-0.00055$\\
$50$ & $5.5$/$0.036$ & $-2.0$/$-2.7$ &
$2.0$/$-2.7$ & $5.45\times 10^{-6}$ & $2.56\times 10^{-6}$
\end{tabular}
\end{table}

One can see from the table that in the case B, i. e.,
the large $\tan\beta$ case,
Br($B_s \rightarrow \mu^+ \mu^-$) can be enhanced by a factor of $10^3$,
compared to SM.
But in the case A where $\tan\beta$ is small, Br($B_s \rightarrow \mu^+ \mu^-$)
is the same order as that in SM. These two case can be discriminated
at Tevatron Run II. If the observed
Br($B_s \rightarrow \mu^+ \mu^-$) is larger than the Standard Model
expectation value by a factor of $10$ or larger and
one assumes that new physics is SUSY, then
this will unambiguously signal the large $\tan\beta$ case.
As for the semileptonic decay
Br($B\rightarrow X_s \mu^+ \mu^-$), there is a $50\%$ enhancement
for the values of the set of
parameters in the large $\tan\beta$ case, which is closer to the central
value of the experiment result, (\ref{semi}), than SM.

\section{Conclusions}\label{sect:conclusion}
We have examined pure leptonic and semileptonic rare B decays,
$B_s \rightarrow \mu^+ \mu^-$ and $B \rightarrow X_s \mu^+ \mu^-$,
under Br($B \rightarrow X_s \gamma$) and other experimental constraints
in the minimal supersymmetric Standard Model.
In particular, we have in detail analyzed
the dependence of the relevant Wilson coefficients on
new flavor violation parameters, off-diagonal scalar quark mass matrix elements.
We find that under all the relevant experimental constraints,
if only the gluino and SM contributions
are included and assuming large mixing of left-handed and right-handed sbottom,
$C^{(\prime)}_{10}$ can be enhanced by a factor of $10\%$ in two cases,
the heavy gluino
and fine-tuning between $\delta^{dLL}_{23}$($\delta^{dRR}_{23}$)
and $\delta^{dLR}_{23}$($\delta^{dRL}_{23}$). In particular,
it is found that $C^{(\prime)}_{10}$
can be enhanced by $40\%$ compared with SM in the fine-tuning case
under all experimental constraints except for
that from $B\rightarrow X_s g$ and hadronic charmless $B$ decays.
When all the contributions are included and all experimental constraints
are taken into account,
$C^{(\prime)}_{10}$ can be enhanced at most by $20\%$ compared with SM.
When CP-odd Higgs $A^0$ is not too heavy ($\sim 250$GeV),
and $\tan\beta$ is large ($\sim 40$),
neutral Higgs boson penguins with gluino and down-type squark
in the loop can significantly
contribute to the $bs$ transition and the contributions can compete with
those due to the chargino and up-type squark loop.
Comparing with the constrained MSSM,
the Wilson coefficient $C_{10}$ can reach a
larger value due to the gluino and neutralino contributions,
but the largest value of $C_{Q_{1,2}}$
allowed by all the experimental constraints is of the same oder.
From the above results, the following conclusions can be drawn:\\
{\bf A}. Although a $20\%$ enhancement of the Wilson coefficient $C_{10}$,
compared to SM, which is twice of that in CMSSM can be reached in MSSM,
it alone is far from the explanation of data
if Br($B_s \rightarrow \mu^+ \mu^-$) = $2\times 10^{-8}$
will be observed at Tevatron run II. \\
{\bf B}. $C_{Q_{1,2}}$ can reach order of one and order of 0.01
in the large and small $\tan\beta$ case respectively and consequently can lead
to an enhancement of Br($B_s \rightarrow \mu^+ \mu^-$)
by a factor of $10^1-10^3$ in the large $\tan\beta$ case.
Therefore, if Br ($B_s \rightarrow \mu^+ \mu^-) \geq 10^{-8}$ is observed,
there should be new physics and $\tan\beta$ must
be large if new physics is the MSSM.

\section*{Acknowledgments}
One of the authors X.H.W. would like to thank Dr. Junjie Cao, Dr. Wei Liao,
Dr. Qi-Shu Yan and Dr. Yu-Feng Zhou for their generous discussions.
The work was supported in part by the National Nature Science
Foundation of China.

\appendix
\section*{\bf Appendix}\nonumber
Wilson coefficients
$C_i(m_w)$, $C_i^\prime(m_w)$ ($i = 7$, $8$, $9$, $10$, $Q_{1,2}$)
of gluino contribution are as follow
\begin{eqnarray}
C_7^{\tilde{\chi}^\pm}(m_w) &=& - \frac{1}{72\lambda_t} \CXgqsLali
\frac{m_w^2}{\mcxi^2} (\CXgqbLali (3f_1(\sqalcxi) + 2f_2(\sqalcxi)) +
8 \CXgqbRali \frac{\mcxi}{m_b} (3f_3(\sqalcxi) + f_4(\sqalcxi))) \nonumber\\
C_7^{\tilde{\chi}^0}(m_w) &=& \frac{1}{72\lambda_t} \NXgqsLali
\frac{m_w^2}{\mnxi^2} (\NXgqbLali f_2(\sqalnxi) + 4\NXgqbRali
\frac{\mnxi}{m_b} f_4(\sqalnxi) ) \nonumber\\
C_7^{\tilde{\chi}^0\prime}(m_w) &=& \frac{1}{72\lambda_t} \NXgqsRali
\frac{m_w^2}{\mnxi^2} (\NXgqbRali f_2(\sqalnxi) + 4\NXgqbLali
\frac{\mnxi}{m_b} f_4(\sqalnxi) ) \nonumber\\
C_7^{\tilde{g}}(m_w) &=& \frac{1}{27\lambda_t} \frac{g_s^2}{g^2} \GGgqsLal
\frac{m_w^2}{\mgg^2} (\GGgqbLal f_2(\sqalgg) + 4\GGgqbRal
\frac{\mgg}{m_b} f_4(\sqalgg) ) \nonumber\\
C_7^{\tilde{g}\prime}(m_w) &=& \frac{1}{27\lambda_t} \frac{g_s^2}{g^2}\GGgqsRal
\frac{m_w^2}{\mgg^2} (\GGgqbRal f_2(\sqalgg) + 4\GGgqbLal
\frac{\mgg}{m_b} f_4(\sqalgg) ) \nonumber\\
C_8^{\tilde{\chi}^\pm}(m_w) &=& - \frac{1}{24\lambda_t} \CXgqsLali
\frac{m_w^2}{\mcxi^2} (\CXgqbLali f_2(\sqalcxi) + 4\CXgqbRali
\frac{\mcxi}{m_b} f_4(\sqalcxi) ) \nonumber\\
C_8^{\tilde{\chi}^0}(m_w) &=& - \frac{1}{24\lambda_t} \NXgqsLali
\frac{m_w^2}{\mnxi^2} (\NXgqbLali f_2(\sqalnxi) + 4\NXgqbRali
\frac{\mnxi}{m_b} f_4(\sqalnxi) ) \nonumber\\
C_8^{\tilde{\chi}^0\prime}(m_w) &=& - \frac{1}{24\lambda_t} \NXgqsRali
\frac{m_w^2}{\mnxi^2} (\NXgqbRali f_2(\sqalnxi) + 4\NXgqbLali
\frac{\mnxi}{m_b} f_4(\sqalnxi) ) \nonumber\\
C_8^{\tilde{g}}(m_w) &=& \frac{1}{72\lambda_t} \frac{g_s^2}{g^2} \GGgqsLal
\frac{m_w^2}{\mgg^2} (\GGgqbLal (9f_1(\sqalgg) + f_2(\sqalgg)) +
8 \GGgqbRal \frac{\mgg}{m_b} (9f_3(\sqalgg) + \frac{1}{2}f_4(\sqalgg)))
\nonumber\\
C_8^{\tilde{g}\prime}(m_w) &=& \frac{1}{72\lambda_t} \frac{g_s^2}{g^2}
\GGgqsRal \frac{m_w^2}{\mgg^2} (\GGgqbRal (9f_1(\sqalgg) + f_2(\sqalgg)) +
8 \GGgqbLal \frac{\mgg}{m_b} (9f_3(\sqalgg) + \frac{1}{2}f_4(\sqalgg)))\\
C_{9,\gamma}^{\tilde{\chi}^\pm} &=& - \frac{1}{36\lambda_t}
\CXgqbLali \CXgqsLali \frac{m_w^2}{\mcxi^2}
[9f_5(\sqalcxi) - 2f_6(\sqalcxi)] \nonumber\\
C_{9,\gamma}^{\tilde{\chi}^0} &=& - \frac{1}{36\lambda_t}
\NXgqbLali \NXgqsLali \frac{m_w^2}{\mnxi^2} f_6(\sqalnxi) \nonumber\\
C_{9,\gamma}^{\tilde{\chi}^0\prime} &=& - \frac{1}{36\lambda_t}
\NXgqbRali \NXgqsRali \frac{m_w^2}{\mnxi^2} f_6(\sqalnxi) \nonumber\\
C_{9,\gamma}^{\tilde{g}} &=& - \frac{2}{27\lambda_t} \frac{g_s^2}{g^2}
\GGgqbLal \GGgqsLal \frac{m_w^2}{\mgg^2} f_6(\sqalgg) \nonumber\\
C_{9,\gamma}^{\tilde{g}\prime} &=& - \frac{2}{27\lambda_t} \frac{g_s^2}{g^2}
\GGgqbRal \GGgqsRal \frac{m_w^2}{\mgg^2} f_6(\sqalgg) \nonumber\\
C_{9,z}^{\tilde{\chi}^\pm} &=& \frac{1}{16\lambda_t s^2_w} (-1 + 4s_w^2)
\CXgqsLbej \CXgqbLali [ -3 \delta_{ij} (\Gamma^{uL}_{\alpha m}
\Gamma^{uL\ast}_{\beta m}) f_{c00}(\sqalcxj,\sqbecxj) \nonumber\\
&&+ \delta_{\alpha\beta} (-2 (U_{i1} U^\ast_{j1}) \frac{\mcxi}{\mcxj}
f_{c0}(\sqalcxj,\cxicxj)
+ 3 (V^\ast_{i1} V_{j1}) f_{c00}(\sqalcxj,\cxicxj)) ] \nonumber\\
C_{9,z}^{\tilde{\chi}^0} &=& \frac{1}{16\lambda_t s^2_w} (-1 + 4s_w^2)
\NXgqsLbej \NXgqbLali [ 3 \delta_{ij} (\Gamma^{dL}_{\alpha m}
\Gamma^{dL\ast}_{\beta m}) f_{c00}(\sqalnxi,\sqbenxj) \nonumber\\
&&+ \delta_{\alpha\beta} (4 \NXzggLij \frac{\mnxi}{\mnxj}
f_{c0}(\sqalnxj,\nxinxj) - 6 \NXzggRij f_{c00}(\sqalnxj,\nxinxj)) ] \nonumber\\
C_{9,z}^{\tilde{\chi}^0\prime} &=& \frac{1}{16\lambda_t s^2_w} (-1 + 4s_w^2)
\NXgqsRbej \NXgqbRali
[ -3 \delta_{ij} (\Gamma^{dR}_{\alpha m} \Gamma^{dR\ast}_{\beta m})
f_{c00}(\sqalnxi,\sqbenxj) \nonumber\\
&&+ \delta_{\alpha\beta} (4 \NXzggRij \frac{\mnxi}{\mnxj}
f_{c0}(\sqalnxj,\nxinxj) - 6 \NXzggLij f_{c00}(\sqalnxj,\nxinxj)) ] \nonumber\\
C_{9,z}^{\tilde{g}} &=& \frac{1}{6\lambda_t s^2_w} \frac{g^2_s}{g^2}
(-1 + 4s_w^2)
\GGgqsLbe \GGgqbLal [ 3 (\Gamma^{dL}_{\alpha m} \Gamma^{dL\ast}_{\beta m})
f_{c00}(\sqalgg,\sqbegg) ] \nonumber\\
C_{9,z}^{\tilde{g}\prime} &=& \frac{1}{6\lambda_t s^2_w} \frac{g^2_s}{g^2}
(-1 + 4s_w^2)
\GGgqsRbe \GGgqbRal [ -3 (\Gamma^{dR}_{\alpha m} \Gamma^{dR\ast}_{\beta m})
f_{c00}(\sqalgg,\sqbegg) ] \nonumber\\
C^{\tilde{\chi}^\pm}_{9, {\rm box}} &=& \frac{1}{2\lambda_t s^2_w}
\CXgqsLali \CXgqbLbej [ -\frac{1}{6} \delta_{\alpha\beta} \frac{m_w^2}{\mcxj^2}
\CXgnuLstari \CXgnuLj f_{d00}(\sqalcxj, \cxicxj, \snucxj)] \nonumber\\
C^{\tilde{\chi}^0}_{9, {\rm box}} &=& \frac{1}{2\lambda_t s^2_w}
\NXgqsLali \NXgqbLbej [ -\frac{1}{6} \delta_{\alpha\beta} \frac{m_w^2}{\mnxj^2}
(\frac{\mnxi}{\mnxj} (\NXglRstarim \NXglRjm - \NXglLim \NXglLstarjm)
f_{d0}(\sqalnxj, \nxinxj, \slmj)\nonumber\\
&& + (\NXglLstarim \NXglLjm - \NXglRim \NXglRstarjm)
f_{d00}(\sqalnxj, \nxinxj, \slmj))] \nonumber\\
C^{\tilde{\chi}^0\prime}_{9, {\rm box}} &=& \frac{1}{2\lambda_t s^2_w}
\NXgqsRali \NXgqbRbej [- \frac{1}{6} \delta_{\alpha\beta} \frac{m_w^2}{\mnxj^2}
(\frac{\mnxi}{\mnxj} (\NXglLstarim \NXglLjm - \NXglRim \NXglRstarjm)
f_{d0}(\sqalnxj, \nxinxj, \slmj)\nonumber\\
&& + (\NXglRstarim \NXglRjm - \NXglLim \NXglLstarjm)
f_{d00}(\sqalnxj, \nxinxj, \slmj))] \nonumber\\
C_{10,z}^{\tilde{\chi}^\pm} &=& \frac{-1}{1 - 4s_w^2}
C_{9,z}^{\tilde{\chi}^\pm} \nonumber\\
C_{10,z}^{\tilde{\chi}^0} &=& \frac{-1}{1 - 4s_w^2}
C_{9,z}^{\tilde{\chi}^0} \nonumber\\
C_{10,z}^{\tilde{\chi}^0\prime} &=& \frac{-1}{1 - 4s_w^2}
C_{9,z}^{\tilde{\chi}^0\prime} \nonumber\\
C_{10,z}^{\tilde{g}} &=& \frac{-1}{1 - 4s_w^2}
C_{9,z}^{\tilde{g}} \nonumber\\
C_{10,z}^{\tilde{g}\prime} &=& \frac{-1}{1 - 4s_w^2}
C_{10,z}^{\tilde{g}\prime} \nonumber\\
C^{\tilde{\chi}^\pm}_{10, {\rm box}} &=& -
C^{\tilde{\chi}^\pm}_{9, {\rm box}} \nonumber\\
C^{\tilde{\chi}^0}_{10, {\rm box}} &=& \frac{1}{2\lambda_t s^2_w}
\NXgqsLali \NXgqbLbej [ \frac{1}{6} \delta_{\alpha\beta} \frac{m_w^2}{\mnxj^2}
(- \frac{\mnxi}{\mnxj} (\NXglRstarim \NXglRjm + \NXglLim \NXglLstarjm)
f_{d0}(\sqalnxj, \nxinxj, \slmj)\nonumber\\
&& + (\NXglLstarim \NXglLjm + \NXglRim \NXglRstarjm)
f_{d00}(\sqalnxj, \nxinxj, \slmj))] \nonumber\\
C^{\tilde{\chi}^0\prime}_{10, {\rm box}} &=& \frac{1}{2\lambda_t s^2_w}
\NXgqsRali \NXgqbRbej [ \frac{1}{6} \delta_{\alpha\beta} \frac{m_w^2}{\mnxj^2}
(- \frac{\mnxi}{\mnxj} (\NXglLstarim \NXglLjm + \NXglRim \NXglRstarjm)
f_{d0}(\sqalnxj, \nxinxj, \slmj)\nonumber\\
&& + (\NXglRstarim \NXglRjm + \NXglLim \NXglLstarjm)
f_{d00}(\sqalnxj, \nxinxj, \slmj))] \\
C^{\tilde{\chi}^\pm}_{Q_1} &=& \frac{1}{2\lambda_t s^2_w} \CXgqbRbej \CXgqsLali
[ \delta_{ij} \delta_{\alpha\beta} \frac{m_l \mcxj}{m^2_{H^0}}
\frac{1}{\cos^2\beta} (\cos^2\alpha + r_s\sin^2\alpha) f_{b0}(\sqalcxj)
\nonumber\\
&& + \delta_{ij} \frac{m_l m_w}{m^2_{H^0}} \frac{1}{m_{\tilde{\chi}^-_j}}
\frac{1}{\cos\beta}
(- \shHualbe \cos\alpha + r_s \shhualbe \sin\alpha)
f_{c0}(\sqalcxj, \sqbecxj) \nonumber\\
&& + \delta_{\alpha\beta} \frac{m_l m_w}{m^2_{H^0}} \frac{1}{\cos\beta}
( \frac{\mcxi}{\mcxj} (-\gcxHRij \cos\alpha + r_s\gcxhRij\sin\alpha)
f_{c0}(\sqalcxj, \cxicxj) \nonumber\\
&&+ 3(- \gcxHLij \cos\alpha + r_s\gcxhLij \sin\alpha)
f_{c00}(\sqalcxj, \cxicxj))
] \nonumber\\
C^{\tilde{\chi}^\pm}_{\rm Q_{1}, box} &=& \frac{1}{2\lambda_t s^2_w}
\CXgqbRbej \CXgqsLali
[ \frac{1}{6} \frac{m_w^2}{\mcxj^2} (- \frac{\mcxi}{\mcxj}
\CXgnuLstari \CXgnuRj f_{d0}(\sqalcxj, \cxicxj, \snucxj) \nonumber\\
&&+ 2\CXgnuRstari \CXgnuLj f_{d00}(\sqalcxj, \cxicxj, \snucxj))
] \nonumber\\
C^{\tilde{\chi}^0}_{Q_{1}} &=& \frac{1}{2\lambda_t s^2_w} \NXgqbRbej \NXgqsLali
[ \delta_{ij} \delta_{\alpha\beta} \frac{m_l \mnxj}{m^2_{H^0}}
\frac{1}{\cos^2\beta} (\cos^2\alpha + r_s\sin^2\alpha) f_{b0}(\sqalnxj)
\nonumber\\
&& + \delta_{ij} \frac{m_l m_w}{m^2_{H^0}} \frac{1}{m_{\tilde{\chi}^0_j}}
\frac{1}{\cos\beta} (- \shHdalbe \cos\alpha + r_s \shhdalbe \sin\alpha)
f_{c0}(\sqalnxj, \sqbenxj) \nonumber\\
&& + \delta_{\alpha\beta} \frac{m_l m_w}{m^2_{H^0}} \frac{1}{\cos\beta}
( \frac{\mnxi}{\mnxj} (-\gnxHRij \cos\alpha + r_s \gnxhRij\sin\alpha)
f_{c0}(\sqalnxj, \nxinxj) \nonumber\\
&&+ 3(- \gnxHLij \cos\alpha + r_s \gnxhLij \sin\alpha)
f_{c00}(\sqalnxj, \nxinxj))
] \nonumber\\
C^{\tilde{\chi}^0}_{\rm Q_{1}, box} &=& \frac{1}{2\lambda_t s^2_w}
\NXgqbRbej \NXgqsLali
[ \frac{1}{6} \frac{m_w^2}{\mnxj^2} (- \frac{\mnxi}{\mnxj}
(\NXglLstarim \NXglRjm - \NXglRim \NXglLstarjm)
f_{d0}(\sqalnxj, \nxinxj, \slmnxj) \nonumber\\
&&+ 2(\NXglRstarim \NXglLjm - \NXglLim \NXglRstarjm)
f_{d00}(\sqalnxj, \nxinxj, \slmnxj))
] \nonumber\\
C^{\tilde{\chi}^0\prime}_{Q_{1}} &=& \frac{1}{2\lambda_t s^2_w} \NXgqbLbej \NXgqsRali
[ \delta_{ij} \delta_{\alpha\beta} \frac{m_l \mnxj}{m^2_{H^0}}
\frac{1}{\cos^2\beta} (\cos^2\alpha + r_s \sin^2\alpha) f_{b0}(\sqalnxj)
\nonumber\\
&& + \delta_{ij} \frac{m_l m_w}{m^2_{H^0}} \frac{1}{m_{\tilde{\chi}^0_j}}
\frac{1}{\cos\beta} (- \shHdalbe \cos\alpha + r_s \shhdalbe \sin\alpha)
f_{c0}(\sqalnxj, \sqbenxj) \nonumber\\
&& + \delta_{\alpha\beta} \frac{m_l m_w}{m^2_{H^0}} \frac{1}{\cos\beta}
( \frac{\mnxi}{\mnxj} (-\gnxHLij \cos\alpha + r_s \gnxhLij\sin\alpha)
f_{c0}(\sqalnxj, \nxinxj) \nonumber\\
&&+ 3(- \gnxHRij \cos\alpha + r_s \gnxhRij \sin\alpha)
f_{c00}(\sqalnxj, \nxinxj))
] \nonumber\\
C^{\tilde{\chi}^0\prime}_{\rm Q_{1}, box} &=& \frac{1}{2\lambda_t s^2_w}
\NXgqbLbej \NXgqsRali
[ \frac{1}{6} \frac{m_w^2}{\mnxj^2} (- \frac{\mnxi}{\mnxj}
(\NXglRstarim \NXglLjm - \NXglLim \NXglRstarjm)
f_{d0}(\sqalnxj, \nxinxj, \slmnxj) \nonumber\\
&&+ 2(\NXglLstarim \NXglRjm - \NXglRim \NXglLstarjm)
f_{d00}(\sqalnxj, \nxinxj, \slmnxj))
] \nonumber\\
C^{\tilde{g}}_{Q_{1}} &=& \frac{4}{3\lambda_t s^2_w} \frac{g^2_s}{g^2}
\GGgqbRbe \GGgqsLal
[ \delta_{\alpha\beta} \frac{m_l \mgg}{m^2_{H^0}}
\frac{1}{\cos^2\beta} (\cos^2\alpha + r_s \sin^2\alpha) f_{b0}(\sqalgg)
\nonumber\\
&& + \frac{m_l m_w}{m^2_{H^0}} \frac{1}{m_{\tilde{g}}}
\frac{1}{\cos\beta} (- \shHdalbe \cos\alpha + r_s \shhdalbe \sin\alpha)
f_{c0}(\sqalgg, \sqbegg)
] \nonumber\\
C^{\tilde{g}\prime}_{Q_{1}} &=& \frac{4}{3\lambda_t s^2_w} \frac{g^2_s}{g^2}
\GGgqbLbe \GGgqsRal
[ \delta_{\alpha\beta} \frac{m_l \mgg}{m^2_{H^0}}
\frac{1}{\cos^2\beta} (\cos^2\alpha + r_s \sin^2\alpha) f_{b0}(\sqalgg)
\nonumber\\
&& + \frac{m_l m_w}{m^2_{H^0}}  \frac{1}{m_{\tilde{g}}} \frac{1}{\cos\beta}
(- \shHdalbe \cos\alpha + r_s \shhdalbe \sin\alpha)
f_{c0}(\sqalgg, \sqbegg)
] \nonumber\\
C^{\tilde{\chi}^\pm}_{Q_2} &=& \frac{1}{2\lambda_t s^2_w} \CXgqbRbej \CXgqsLali
[ - \delta_{ij} \delta_{\alpha\beta} \frac{m_l \mcxj}{m^2_{A^0}}
(r_p + \tan^2\beta) f_{b0}(\sqalcxj)\nonumber\\
&& - \delta_{ij} \frac{m_l m_w}{m^2_{A^0}}
(r_p \shGualbe - \shAualbe \tan\beta)
f_{c0}(\sqalcxj, \sqbecxj) \nonumber\\
&& - \delta_{\alpha\beta} \frac{m_l m_w}{m^2_{A^0}}
( \frac{\mcxi}{\mcxj} (r_p \gcxGRij - \gcxARij\tan\beta)
f_{c0}(\sqalcxj, \cxicxj) \nonumber\\
&&+ 3(r_p \gcxGLij - \gcxALij \tan\beta) f_{c00}(\sqalcxj, \cxicxj))
] \nonumber\\
C^{\tilde{\chi}^\pm}_{\rm Q_2, box} &=& \frac{1}{2\lambda_t s^2_w}
\CXgqbRbej \CXgqsLali
[ - \frac{1}{6} \frac{m_w^2}{\mcxj^2} (\frac{\mcxi}{\mcxj}
\CXgnuLstari \CXgnuRj f_{d0}(\sqalcxj, \cxicxj, \snucxj) \nonumber\\
&&+ 2\CXgnuRstari \CXgnuLj f_{d00}(\sqalcxj, \cxicxj, \snucxj))
] \nonumber\\
C^{\tilde{\chi}^0}_{Q_2} &=& \frac{1}{2\lambda_t s^2_w} \NXgqbRbej \NXgqsLali
[- \delta_{ij} \delta_{\alpha\beta} \frac{m_l \mnxj}{m^2_{A^0}}
(r_p + \tan^2\beta) f_{b0}(\sqalnxj)\nonumber\\
&& - \delta_{ij} \frac{m_l m_w}{m^2_{A^0}}  \frac{1}{m_{\tilde{\chi}^0_j}}
(r_p \shGdalbe - \shAdalbe \tan\beta)
f_{c0}(\sqalnxj, \sqbenxj) \nonumber\\
&& - \delta_{\alpha\beta} \frac{m_l m_w}{m^2_{A^0}}
( \frac{\mnxi}{\mnxj} (r_p \gnxGRij - \gnxARij\tan\beta)
f_{c0}(\sqalnxj, \nxinxj) \nonumber\\
&&+ 3(r_p \gnxGLij - \gnxALij \tan\beta)
f_{c00}(\sqalnxj, \nxinxj))
] \nonumber\\
C^{\tilde{\chi}^0}_{\rm Q_2, box} &=& \frac{1}{2\lambda_t s^2_w}
\NXgqbRbej \NXgqsLali
[ - \frac{1}{6} \frac{m_w^2}{\mnxj^2} (\frac{\mnxi}{\mnxj}
(\NXglLstarim \NXglRjm - \NXglLim \NXglRstarjm)
f_{d0}(\sqalnxj, \nxinxj, \slmnxj) \nonumber\\
&&+ 2(\NXglRstarim \NXglLjm - \NXglLim \NXglRstarjm)
f_{d00}(\sqalnxj, \nxinxj, \slmnxj))
] \nonumber\\
C^{\tilde{\chi}^0\prime}_{Q_2} &=& \frac{1}{2\lambda_t s^2_w} \NXgqbLbej \NXgqsRali
[\delta_{ij} \delta_{\alpha\beta} \frac{m_l \mnxj}{m^2_{A^0}}
(r_p + \tan^2\beta) f_{b0}(\sqalnxj)\nonumber\\
&& - \delta_{ij} \frac{m_l m_w}{m^2_{A^0}} \frac{1}{m_{\tilde{\chi}^0_j}}
(r_p \shGdalbe - \shAdalbe \tan\beta)
f_{c0}(\sqalnxj, \sqbenxj) \nonumber\\
&& - \delta_{\alpha\beta} \frac{m_l m_w}{m^2_{A^0}}
( \frac{\mnxi}{\mnxj} (r_p \gnxGLij - \gnxALij\tan\beta)
f_{c0}(\sqalnxj, \nxinxj) \nonumber\\
&&+ 3(r_p \gnxGRij - \gnxARij \tan\beta)
f_{c00}(\sqalnxj, \nxinxj))
] \nonumber\\
C^{\tilde{\chi}^0\prime}_{\rm Q_2, box} &=& \frac{1}{2\lambda_t s^2_w}
\NXgqbLbej \NXgqsRali
[\frac{1}{6} \frac{m_w^2}{\mnxj^2} (\frac{\mnxi}{\mnxj}
(\NXglRstarim \NXglLjm - \NXglLim \NXglRstarjm)
f_{d0}(\sqalnxj, \nxinxj, \slmnxj) \nonumber\\
&&+ 2(\NXglLstarim \NXglRjm - \NXglRim \NXglLstarjm)
f_{d00}(\sqalnxj, \nxinxj, \slmnxj))
] \nonumber\\
C^{\tilde{g}}_{Q_2} &=& \frac{4}{3\lambda_t s^2_w} \frac{g^2_s}{g^2}
\GGgqbRbe \GGgqsLal
[ - \delta_{\alpha\beta} \frac{m_l \mgg}{m^2_{A^0}}
(r_p + \tan^2\beta) f_{b0}(\sqalgg)\nonumber\\
&& - \frac{m_l m_w}{m^2_{A^0}} \frac{1}{m_{\tilde{g}}}
(r_p \shGdalbe - \shAdalbe \tan\beta)
f_{c0}(\sqalgg, \sqbegg)
] \nonumber\\
C^{\tilde{g}\prime}_{Q_2} &=& \frac{4}{3\lambda_t s^2_w} \frac{g^2_s}{g^2}
\GGgqbLbe \GGgqsRal
[ \delta_{\alpha\beta} \frac{m_l \mgg}{m^2_{A^0}}
(r_p + \tan^2\beta) f_{b0}(\sqalgg)\nonumber\\
&& - \frac{m_l m_w}{m^2_{A^0}} \frac{1}{m_{\tilde{g}}}
(r_p \shGdalbe - \shAdalbe \tan\beta)
f_{c0}(\sqalgg, \sqbegg)
]
\end{eqnarray}
where $r_s = \frac{m^2_{H^0}}{m^2_{h^0}}$,
$r_p = \frac{m^2_{A^0}}{m^2_{Z^0}}$, and $x_{ij} = m^2_i/m^2_j$,
$s_w = \sin\theta_w$, $c_w = \cos\theta_w$.
We have check $C_{7, 8, 9, 10}$ with ref.~\cite{goto97}, our results agree with
them except there is a minus sign difference with their box diagram.
$C_{Q_{1,2}}$ calculated here agree with ref.~\cite{urban}.

The one-loop functions are normalized to be $1$, if all the arguments are
equal and set to be $1$.
\begin{eqnarray}
f_1(x)&=& 2(1 - 6x + 3x^2 + 2x^3 - 6x^2\ln(x))/(1 - x)^4 \nonumber\\
f_2(x)&=& 2(2 + 3x - 6x^2 + x^3 + 6x\ln(x))/(1 - x)^4 \nonumber\\
f_3(x)&=& 3(1 - 4x + 3x^2 - 2x^2\ln(x))/(2(1 - x)^3) \nonumber\\
f_4(x)&=& 3(1 - x^2 + 2x\ln(x))/(1 - x)^3 \nonumber\\
f_5(x)&=& 2(7 - 36x + 45x^2 - 16x^3 + 6x^2(-3 + 2x)\ln(x))/(9(1 - x)^4)
\nonumber\\
f_6(x)&=& 2(-11 + 18x - 9x^2 + 2x^3 - 6\ln(x))/(3(1 - x)^4)\nonumber\\
f_{b0}(x) &=& - \frac{x\ln x}{1-x}\nonumber\\
f_{c0}(x,y) &=& -2 [ \frac{x\ln x}{(1-x)(x-y)} + \frac{y\ln y}{(1-y)(y-x)} ]
\nonumber\\
f_{c00}(x,y) &=& - \frac{2}{3} [\frac{x^2\ln x}{(1-x)(x-y)} +
\frac{y^2\ln y}{(1-y)(y-x)} ]\nonumber\\
f_{d0}(x,y,z) &=& 6 [ \frac{x\ln x}{(1-x)(x-y)(x-z)} +
\frac{y\ln y}{(1-y)(y-x)(y-z)} + \frac{z\ln z}{(1-z)(z-x)(z-y)} ]\nonumber\\
f_{d00}(x,y,z) &=& -3 [ \frac{x^2\ln x}{(1-x)(x-y)(x-z)} +
\frac{y^2\ln y}{(1-y)(y-x)(y-z)} + \frac{z^2\ln z}{(1-z)(z-x)(z-y)} ]
\end{eqnarray}

We follow the convension of Haber and Kane~\cite{haberkane}, to present our
Feynman rules and mass matrices of chargino, neutralino, squark, and sleptons.
The chargino mass matrix $X$ is diagonalized by two matrix $U$ and $V$, with
$U^\ast X V^{-1} = {\rm diag}(m_{\tilde{\chi}^\pm_1}, m_{\tilde{\chi}^\pm_2)}$.
The neutralino mass matrix $Y$ is diagonalized by matrix $N$,
$N^\ast Y N^{-1} =
{\rm diag}(m_{\tilde{\chi}^0_1}, m_{\tilde{\chi}^0_2},
m_{\tilde{\chi}^0_3}, m_{\tilde{\chi}^0_4})$,
where $m_{\tilde{\chi}^0_i}(i=1,2,3,4)$ are positive.

The chargino(gluino, neutralino)-squark-quark couplings are as follow
\begin{eqnarray}
{\cal L} &=& g \sum_{i=1,\alpha=1}^{2,6} \bar{\tilde{\chi}^-_i}
 \tilde{u}^\dagger_\alpha (G^{Ld}_{\tilde{u}_\alpha \tilde{\chi}_i^-} P_L +
 G^{Rd}_{\tilde{u}_\alpha \tilde{\chi}_i^-} P_R) d +
 g \sum_{i=1,\alpha=1}^{4,6} \bar{\tilde{\chi}^0_i}
 \tilde{d}^\dagger_\alpha (G^{Ld}_{\tilde{d}_\alpha \tilde{\chi}_i^0} P_L +
 G^{Rd}_{\tilde{d}_\alpha \tilde{\chi}_i^0} P_R) d \nonumber\\
 && + \sqrt{2} g_s \sum_{\alpha=1}^{6} \bar{\tilde{g}}
 \tilde{d}^\dagger_\alpha T^a (G^{Ld}_{\tilde{d}_\alpha \tilde{g}} P_L +
 G^{Rd}_{\tilde{d}_\alpha \tilde{g}} P_R) d + {\rm H.C.}\nonumber\\
G^{Ls}_{\tilde{u}_\alpha \tilde{\chi}_i^-} &=& (- V^\ast_{i1}
 \Gamma^{uL}_{\alpha m} + V^\ast_{i2} \Gamma^{uR}_{\alpha m}
 \frac{m_{u_m}}{\sqrt{2} m_w s_\beta} ) K_{m2}, \hspace{5mm}
G^{Rs}_{\tilde{u}_\alpha \tilde{\chi}_i^-} = 0 \nonumber\\
G^{Lb}_{\tilde{u}_\alpha \tilde{\chi}_i^-} &=& (- V^\ast_{i1}
 \Gamma^{uL}_{\alpha m} + V^\ast_{m2} \Gamma^{uR}_{\alpha m}
 \frac{m_{u_m}}{\sqrt{2} m_w s_\beta} ) K_{m3}, \hspace{5mm}
G^{Rb}_{\tilde{u}_\alpha \tilde{\chi}_i^-} = U_{i2} \Gamma^{uL}_{\alpha m}
 K_{m3} \frac{m_{b}}{\sqrt{2} m_w c_\beta} \nonumber\\
G^{Ls}_{\tilde{d}_\alpha \tilde{\chi}^0_i} &=&
 \sqrt{2} (- \frac{1}{6} t_w N_{i1}^\ast +
 \frac{1}{2} N_{i2}^\ast) \Gamma^{dL}_{\alpha 2}, \hspace{5mm}
G^{Rs}_{\tilde{d}_\alpha \tilde{\chi}^0_i} =
 - \frac{\sqrt{2}}{3} t_w N_{i1} \Gamma^{dR}_{\alpha 2}\nonumber\\
\NXgqbLali &=& \sqrt{2} (- \frac{1}{6} t_w N_{i1}^\ast +
 \frac{1}{2} N_{i2}^\ast) \Gamma^{dL}_{\alpha 3} -
 N_{i3}^\ast \Gamma^{dR}_{\alpha 3}
 \frac{m_{b}}{\sqrt{2} m_w c_\beta}, \hspace{3mm}
\NXgqbRali = - \frac{\sqrt{2}}{3} t_w N_{i1}
 \Gamma^{dR}_{\alpha 3} - N_{i3} \Gamma^{dL}_{\alpha 3}
 \frac{m_{b}}{\sqrt{2} m_w c_\beta} \nonumber\\
G^{Ls}_{\tilde{d}_\alpha \tilde{g}} &=& - \Gamma^{dL}_{\alpha 2}, \hspace{5mm}
G^{Rs}_{\tilde{d}_\alpha \tilde{g}} =
 \Gamma^{dR}_{\alpha 2}\nonumber\\
\GGgqbLal &=& - \Gamma^{dL}_{\alpha 3}, \hspace{5mm}
\GGgqbRal = \Gamma^{dR}_{\alpha 3}
\end{eqnarray}
The chargino(neutralino)-lepton-slepton couplings are
\begin{eqnarray}
{\cal L} &=& g \sum_{i=1,\alpha=1}^{2,3} \bar{\tilde{\chi}^-_i}
 \tilde{\nu}^\dagger_\alpha G^{Ll}_{\tilde{\nu}_\alpha \tilde{\chi}_i^-}
P_L \mu +
 g \sum_{i=1,\alpha=1}^{4,6} \bar{\tilde{\chi}^0_i}
 \tilde{l}^\dagger_\alpha (G^{Ll}_{\tilde{l}_\alpha \tilde{\chi}_i^0} P_L +
 G^{Rl}_{\tilde{l}_\alpha \tilde{\chi}_i^0} P_R) \mu \nonumber\\
G^{Ll}_{\tilde{\nu}_\alpha \tilde{\chi}_i^-} &=& - V_{i1}^\ast
\Gamma^{\nu L}_{m2} \nonumber\\
G^{Ll}_{\tilde{l}_\alpha \tilde{\chi}_i^0} &=& \sqrt{2}
(\frac{1}{2} t_w N_{i1}^\ast + \frac{1}{2} N_{i2}^\ast )
\Gamma^{lL}_{m2}, \hspace{5mm}
G^{Rl}_{\tilde{l}_\alpha \tilde{\chi}_i^0} = - \sqrt{2} t_w N_{i1}
\Gamma^{lR}_{m2}
\end{eqnarray}
The Z-squark-squark couplings are
\begin{eqnarray}
{\cal L} &=& - \frac{g}{c_w} Z_\mu (\tilde{q}^\ast_\alpha
\stackrel{\leftrightarrow}{\partial^{\mu}} \tilde{q}_\beta)
G^{z}_{\tilde{q}_\alpha \tilde{q}_\beta} \nonumber\\
G^{z}_{\tilde{u}_\alpha \tilde{u}_\beta} &=& \frac{1}{2}
(\Gamma^{uL}_{\alpha m} \Gamma^{uL\ast}_{\beta m})
- \frac{2}{3} s^2_w \delta_{\alpha\beta} \nonumber\\
G^{z}_{\tilde{d}_\alpha \tilde{d}_\beta} &=& - \frac{1}{2}
(\Gamma^{dL}_{\alpha m} \Gamma^{dL\ast}_{\beta m}) + \frac{1}{3} s^2_w
\delta_{\alpha\beta}
\end{eqnarray}
The Z-chargino(neutralino)-chargino(neutralino) couplings are
\begin{eqnarray}
{\cal L} &=& - \frac{g}{c_w} Z_{\mu} [ \bar{\tilde{\chi}^-_i} \gamma^\mu
(G^{Lz}_{\tilde{\chi}^-_i \tilde{\chi}^-_j} P_L +
G^{Rz}_{\tilde{\chi}^-_i \tilde{\chi}^-_j} P_R) \tilde{\chi}^-_j +
\frac{1}{2} \bar{\tilde{\chi}^0_i} \gamma^\mu
(G^{Lz}_{\tilde{\chi}^0_i \tilde{\chi}^0_j} P_L +
G^{Rz}_{\tilde{\chi}^0_i \tilde{\chi}^0_j} P_R) \tilde{\chi}^0_j ] \nonumber\\
G^{Lz}_{\tilde{\chi}^-_i \tilde{\chi}^-_j} &=& - \frac{1}{2} U_{i1}
U^\ast_{j1} + \delta_{ij}(s^2_w - \frac{1}{2}), \hspace{5mm}
G^{Rz}_{\tilde{\chi}^-_i \tilde{\chi}^-_j} = - \frac{1}{2} V_{i1}
V^\ast_{j1} + \delta_{ij}(s^2_w - \frac{1}{2})
\nonumber\\
\NXzggLij &=& \frac{1}{2} N_{i3} N_{j3}^\ast -
     \frac{1}{2} N_{i4} N_{j4}^\ast, \hspace{5mm}
\NXzggRij = - G^{Lz^\ast}_{\tilde{\chi}^0_i \tilde{\chi}^0_j}
\end{eqnarray}
The Higgs-chargino(neutralino)-chargino(neutralino) couplings are
\begin{eqnarray}
{\cal L} &=& - g \bar{\tilde{\chi}^-_i}
[h^0(G^{Lh}_{\tilde{\chi}^-_i \tilde{\chi}^-_j} P_L +
G^{Rh}_{\tilde{\chi}^-_i \tilde{\chi}^-_j} P_R) +
H^0 (G^{LH}_{\tilde{\chi}^-_i \tilde{\chi}^-_j} P_L +
G^{RH}_{\tilde{\chi}^-_i \tilde{\chi}^-_j} P_R)] \tilde{\chi}^-_j \nonumber\\
&& + i g \bar{\tilde{\chi}^-_i}
[G^0(G^{LG}_{\tilde{\chi}^-_i \tilde{\chi}^-_j} P_L +
G^{RG}_{\tilde{\chi}^-_i \tilde{\chi}^-_j} P_R) +
A^0 (G^{LA}_{\tilde{\chi}^-_i \tilde{\chi}^-_j} P_L +
G^{RA}_{\tilde{\chi}^-_i \tilde{\chi}^-_j} P_R)] \tilde{\chi}^-_j \nonumber\\
&& - \frac{1}{2} g \bar{\tilde{\chi}^0_i}
[h^0(G^{Lh}_{\tilde{\chi}^0_i \tilde{\chi}^0_j} P_L +
G^{Rh}_{\tilde{\chi}^0_i \tilde{\chi}^0_j} P_R) +
H^0 (G^{LH}_{\tilde{\chi}^0_i \tilde{\chi}^0_j} P_L +
G^{RH}_{\tilde{\chi}^0_i \tilde{\chi}^0_j} P_R)] \tilde{\chi}^0_j \nonumber\\
&& + i \frac{1}{2} g \bar{\tilde{\chi}^0_i}
[G^0(G^{LG}_{\tilde{\chi}^0_i \tilde{\chi}^0_j} P_L +
G^{RG}_{\tilde{\chi}^0_i \tilde{\chi}^0_j} P_R) +
A^0 (G^{LA}_{\tilde{\chi}^0_i \tilde{\chi}^0_j} P_L +
G^{RA}_{\tilde{\chi}^0_i \tilde{\chi}^0_j} P_R)] \tilde{\chi}^0_j \nonumber\\
\gcxhLij &=& - Q^\ast_{ji} s_\alpha + S^\ast_{ji} c_\alpha, \hspace{5mm}
\gcxhRij = - Q_{ij} s_\alpha + S_{ij} c_\alpha \nonumber\\
\gcxHLij &=& Q^\ast_{ji} c_\alpha + S^\ast_{ji} s_\alpha, \hspace{5mm}
\gcxHRij = Q_{ij} c_\alpha + S_{ij} s_\alpha \nonumber\\
\gcxGLij &=& - Q^\ast_{ji} c_\beta + S^\ast_{ji} s_\beta, \hspace{5mm}
\gcxGRij = Q_{ij} c_\beta - S_{ij} s_\beta \nonumber\\
\gcxALij &=& Q^\ast_{ji} s_\beta + S^\ast_{ji} c_\beta, \hspace{5mm}
\gcxARij = - Q_{ij} s_\beta - S_{ij} c_\beta \nonumber\\
Q_{ij} &=& \sqrt{\frac{1}{2}} V_{i1} U_{j2}, \hspace{5mm}
S_{ij} = \sqrt{\frac{1}{2}} V_{i2} U_{j1} \nonumber\\
\gnxhLij &=& - Q^{\prime\prime\ast}_{ji} s_\alpha -
     S^{\prime\prime\ast}_{ji} c_\alpha, \hspace{5mm}
\gnxhRij = - Q^{\prime\prime}_{ij} s_\alpha -
     S^{\prime\prime}_{ij} c_\alpha \nonumber\\
\gnxHLij &=& Q^{\prime\prime\ast}_{ji} c_\alpha -
     S^{\prime\prime\ast}_{ji} s_\alpha, \hspace{5mm}
\gnxHRij = Q^{\prime\prime}_{ij} c_\alpha -
     S^{\prime\prime}_{ij} s_\alpha \nonumber\\
\gnxGLij &=& - Q^{\prime\prime\ast}_{ji} c_\beta -
     S^{\prime\prime\ast}_{ji} s_\beta \hspace{5mm}
\gnxGRij =  Q^{\prime\prime}_{ij} c_\beta +
     S^{\prime\prime}_{ij} s_\beta\nonumber\\
\gnxALij &=& Q^{\prime\prime\ast}_{ji} s_\beta -
     S^{\prime\prime\ast}_{ji} c_\beta \hspace{5mm}
\gnxARij =  - Q^{\prime\prime}_{ij} s_\beta +
     S^{\prime\prime}_{ij} c_\beta\nonumber\\
Q^{\prime\prime}_{ij} &=& \frac{1}{2} ( N_{i3} (N_{j2} - N_{j1} t_w) +
     N_{j3} (N_{i2} - N_{i1}tw) ) \nonumber\\
S^{\prime\prime}_{ij} &=& \frac{1}{2} ( N_{i4} (N_{j2} - N_{j1} t_w) +
     N_{j4} (N_{i2} - N_{i1} t_w) )
\end{eqnarray}
where $s_\alpha = \sin\alpha$, $c_\alpha = \cos\alpha$.
The Higgs-squark-squark couplings are
\begin{eqnarray}
{\cal L} &=& - g \tilde{q}^\dagger_\alpha [
h^0 G^{h^0}_{\tilde{q}_\alpha \tilde{q}_\beta} +
H^0 G^{H^0}_{\tilde{q}_\alpha \tilde{q}_\beta}
] \tilde{q}_\beta + i g \tilde{q}^\dagger_\alpha [
G^0 G^{G^0}_{\tilde{q}_\alpha \tilde{q}_\beta} +
A^0 G^{A^0}_{\tilde{q}_\alpha \tilde{q}_\beta}
] \tilde{q}_\beta \nonumber\\
\shhualbe &=& \Gamma^{uL\ast}_{\alpha i} (m_{u_i} \mu s_\alpha \delta_{ij} +
  v_u A^u_{ij} c_\alpha ) \Gamma^{uR}_{\beta j}/(2 m_w s_\beta) +
  {\rm h.c.}\nonumber\\
\shHualbe &=& \Gamma^{uL\ast}_{\alpha i} (- m_{u_i} \mu c_\alpha \delta_{ij} +
  v_u A^u_{ij} s_\alpha ) \Gamma^{uR}_{\beta j}/(2 m_w s_\beta) +
  {\rm h.c.}\nonumber\\
\shGualbe &=& \Gamma^{uR\ast}_{\alpha i} (- m_{u_i} \mu \cot\beta \delta_{ij}
  + v_u A^u_{ij} ) \Gamma^{uL}_{\beta j}/(2 m_w) - {\rm h.c.} \nonumber\\
\shAualbe &=& \Gamma^{uR\ast}_{\alpha i} ( m_{u_i} \mu \delta_{ij} +
  v_u A^u_{ij} \cot\beta ) \Gamma^{uL}_{\beta j}/(2 m_w)
  - {\rm h.c.}\nonumber\\
\shhdalbe &=& \Gamma^{dL\ast}_{\alpha i} ( - m_{d_i} \mu c_\alpha \delta_{ij}
  - v_d A^d_{ij} s_\alpha ) \Gamma^{dR}_{\beta j}/(2 m_w c_\beta) +
     {\rm h.c.}\nonumber\\
\shHdalbe &=& \Gamma^{dL\ast}_{\alpha i} ( - m_{d_i} \mu s_\alpha \delta_{ij} +
  v_d A^d_{ij} c_\alpha ) \Gamma^{dR}_{\beta j}/(2 m_w c_\beta) +
     {\rm h.c.}\nonumber\\
\shGdalbe &=& \Gamma^{dR\ast}_{\alpha i} ( m_{d_i} \mu \tan\beta \delta_{ij}
  - v_d A^d_{ij} ) \Gamma^{dL}_{\beta j}/(2 m_w) - {\rm h.c.} \nonumber\\
\shAdalbe &=& \Gamma^{dR\ast}_{\alpha i} ( m_{d_i} \mu \delta_{ij} +
  v_d A^d_{ij} \tan\beta ) \Gamma^{dL}_{\beta j}/(2 m_w) - {\rm h.c.}
\end{eqnarray}

\newpage

\newpage

\newpage

\newpage

\newpage

\end{document}